\documentclass[12pt]{iopart}
\expandafter\let\csname equation*\endcsname\relax
\expandafter\let\csname endequation*\endcsname\relax
\usepackage{amsmath}
\usepackage{amssymb,amsfonts}
\usepackage{bm}
\usepackage{graphicx,epstopdf}
\usepackage{multirow}
\usepackage[caption=false]{subfig}
\usepackage{algorithmic}
\usepackage{algorithm}
\usepackage{adjustbox} 
\usepackage{mathtools}

\usepackage{array} 
\newcolumntype{R}{>{$}r<{$}} %
\newcolumntype{V}[1]{>{[\;}*{#1}{R@{\;\;}}R<{\;]}}
\renewcommand{\vec}[1]{\mathbf{#1}}

\newcommand{\Vf}{\vec{V}}

\newcommand{\vz}{\vec{0}}
\newcommand{\vc}{\vec{c}}
\newcommand{\vd}{\vec{p}}
\newcommand{\vr}{\bm{r}}
\newcommand{\vp}{\ensuremath{\vec{p}_0}}
\newcommand{\vl}{\boldsymbol{\lambda}}

\newcommand{\x}{\vec{x}}
\newcommand{\z}{\vec{z}}

\newcommand{\zpx}{\vec{z}^x_{\vp}}
\newcommand{\zpy}{\vec{z}^y_{\vp}}
\newcommand{\zcx}{\vec{z}^x_{\vc}}
\newcommand{\zcy}{\vec{z}^y_{\vc}}

\newcommand{\Hc}{\vec{H}}
\newcommand{\M}{\vec{M}}
\newcommand{\D}{\vec{D}}
\newcommand{\Dx}{\vec{D}^x}
\newcommand{\Dy}{\vec{D}^y}

\newcommand{\vf}{\vec{v}}
\newcommand{\X}{\mathcal{X}}
\newcommand{\Z}{\mathcal{Z}}
\newcommand{\weip}{w_{\vp}}
\newcommand{\weic}{w_{\vc}}

\newcommand{\argmin}{\mathop{\mathrm{arg\,min}}}

\lineup
\begin{document}

\title[Constrained joint reconstruction for PACT]{Revisiting the joint estimation of initial pressure and speed-of-sound distributions in photoacoustic computed tomography with consideration of canonical object constraints}

\author{Gangwon Jeong$^1$, Umberto Villa$^2$, 
Mark A. Anastasio$^1$\footnote{Author to whom any correspondence should be addressed.}}
\address{$^1$ Department of Bioengineering, University of Illinois Urbana–Champaign, Urbana, IL 61801, United States of America}
\address{$^2$ Oden Institute for Computational Engineering $\&$ Sciences, The University of Texas at Austin, Austin, TX 78712, United States of America}
\ead{maa@illinois.edu}
\vspace{10pt}
\begin{indented}
\item[]August 2024
\end{indented}

\begin{abstract}
In photoacoustic computed tomography (PACT) the accurate estimation of the initial pressure (IP) distribution generally requires knowledge of the object's heterogeneous speed-of-sound (SOS) distribution.
Although hybrid imagers that combine ultrasound tomography with PACT have been proposed, in many current applications of PACT the SOS distribution remains unknown.
Joint reconstruction (JR) of the IP and SOS distributions from PACT measurement data alone can address this issue. 
However, this joint estimation problem is ill-posed and corresponds to a non-convex optimization problem. 
While certain regularization strategies have been deployed, stabilizing the JR problem to yield accurate estimates of the IP and SOS distributions has remained an open challenge.
To address this, the presented numerical studies explore the effectiveness of easy to implement canonical object constraints for stabilizing the JR problem.
The considered constraints include support, bound, and total variation constraints,  which are incorporated into an optimization-based method for JR.
Computer-simulation studies that employ anatomically realistic numerical breast phantoms are conducted to evaluate the impact of these object constraints on JR accuracy.
Additionally, the impact of certain data inconsistencies, such as caused by measurement noise and physics modeling mismatches, on the effectiveness of the object constraints is investigated. 
The results demonstrate, for the first time, that the incorporation of canonical object constraints in an optimization-based image reconstruction method holds significant potential for mitigating the ill-posed nature of the PACT JR problem.
\end{abstract}

%
\vspace{2pc}

\submitto{\IP}
%
%
%

\section{Introduction}
Photoacoustic computed tomography (PACT) is an emerging biomedical imaging modality that is being actively developed for various  applications \cite{poudel2019survey, dogan2019optoacoustic, kruger2010photoacoustic,lin2018single, kruger2013dedicated,li2017single,xia2012whole,ermilov2009development}.
In PACT, an object to-be-imaged is illuminated with a short laser pulse of nanosecond-duration.
The biological tissues within the object absorb the optical energy, leading to a localized and transient increase in temperature. 
This rises in temperature causes rapid thermoelastic expansion, generating an initial pressure (IP) distribution within the tissues through a process known as the photoacoustic effect \cite{wang2017photoacoustic, wang2012biomedical}.
Broad-band ultrasound transducers are then employed to detect these photoacoustic pressure wavefields.
Subsequently, these measurements are processed by use of tomographic reconstruction methods to produce estimates of the object's spatially variant IP distribution. 

Traditional image reconstruction methods for PACT often assume that to-be-imaged object and the background into which it is embedded are described by a single speed-of-sound (SOS) value \cite{xu2005universal, huang2010investigation, wang2012simple}.
However, in biological tissues the SOS distribution is generally heterogeneous, and failing to account for this in the image reconstruction process can lead to a degradation of spatial resolution and artifacts \cite{xu2003effects, xu2003time, yoon2012enhancement, cong2015photoacoustic}.
While a single SOS parameter in a conventional reconstruction method can be tuned to maintain the spatial resolution at certain locations, 
this approach cannot simultaneously bring all structures in the reconstructed IP distribution into focus, especially when variations in the SOS distribution are not weak.
On the other hand, if the heterogeneous SOS distribution of the object is known a \textit{priori}, accurate estimates of the IP distribution can be obtained by use of image reconstruction methods that account for the propagation of the induced photoacoustic wavefields in an acoustically heterogeneous medium \cite{xu2003effects, modgil2010image,treeby2010photoacoustic, huang2013full, poudel2019survey}.
One approach to estimate the heterogeneous SOS distribution is to perform adjunct ultrasound tomography \cite{jin2006thermoacoustic, manohar2007concomitant, jose2012speed, xia2013enhancement, ranjbaran2023quantitative}.
However, in most current applications of PACT the SOS distribution remains unknown.

Motivated by the need to account for SOS variations during image reconstruction, various studies have investigated the concurrent estimation of the IP and SOS distributions using only PACT measurements  \cite{zhang2006reconstruction, jiang2006spatially, stefanov2012instability, kirsch2012simultaneous, oksanen2013photoacoustic, liu2015determining, huang2016joint, matthews2018parameterized, cai2019feature}. 
This is referred to here as the joint reconstruction (JR) problem.
Unfortunately, the JR problem is inherently ill-posed \cite{stefanov2012instability} and corresponds to a non-convex optimization problem.
Huang \textit{et al}. \cite{huang2016joint}  investigated the numerical instability of the JR problem through computer simulations. 
They employed an alternating optimization approach, where the IP and SOS estimates were sequentially updated in each iteration. 
Their findings revealed that the reconstruction of the SOS distribution is particularly sensitive to inaccuracies in the IP estimates due to the non-convex nature of the problem. 
Consequently, distorted SOS estimates were found to lead to further degradation in subsequent IP reconstructions, with errors compounding across iterations.

In attempts to mitigate the ill-posed nature of the JR problem, various regularization methods have been explored in numerical studies.
For example, several studies have sought solutions by minimizing a penalized objective function  \cite{huang2016joint, jiang2006spatially, yuan2006simultaneous}. 
However, the reported studies have been preliminary in nature and evidence of their effectiveness for addressing practical applications is largely lacking. 
An alternative regularization approach involves the exploitation of low-dimensional approximate representations of the SOS distribution \cite{zhang2006reconstruction, zhang2008simultaneous, matthews2018parameterized, cai2019feature}. 
By reducing the number of SOS parameters to be estimated as compared to pixel-wise representations, such methods hold potential for stabilizing JR problem. 
However, establishing useful low-dimensional representations may require significant \textit{a priori} knowledge of the to-be-imaged object's SOS distribution, which may not be readily available without the use of adjunctive imaging. 
There remains an important need for the development of practical and effective regularization strategies that can permit accurate JR and thereby advance applications of PACT.

To address this, the presented work explores the effectiveness of easy to implement canonical object constraints for stabilizing the JR problem.
The considered constraints include support, bound, and total variation (TV) constraints, which are well known and have been widely employed in a variety of tomographic image reconstruction problems.  
However, their effectiveness in the context of JR for PACT has not been systematically investigated.
The object constraints are incorporated into a constrained optimization problem that is designed to yield solutions to the JR problem.
Computer-simulation studies that employ anatomically realistic numerical breast phantoms \cite{li20213, park2023stochastic} are conducted to evaluate the effectiveness of the object constraints. 
The impact of certain types of data inconsistencies is also investigated.

The remainder of this article is organized as follows. 
In Section \ref{back}, the salient PACT imaging physics and discretized forward model are presented.
The definition of the considered canonical object constraints,  formulation of the constrained JR problem, and image reconstruction method are described in Section \ref{method}.
The virtual imaging studies and associated numerical results are provided in Sections \ref{sec:ns} and  \ref{results}. 
Finally, the article concludes with a summary of the key findings and topics for future investigation in Section \ref{sec:conclusions}.

\section{Background} \label{back}
\subsection{Notation}
Italic lowercase letters denote scalar-valued functions (e.g., $v$), while bold italic lowercase letters represent vector-valued functions (e.g., $\bm{v}$).
Bold roman lowercase and uppercase letters are used for vectors (e.g., $\vf$) and matrices (e.g., $\Vf$), respectively.
For a vector $\vf$, the element at index $k$ is denoted by $\vf[k]$. 
For a matrix $\Vf$, the element at the $i$-th row and $j$-th column is represented by $\Vf[i,j]$. Additionally, $\Vf[i,:]$ refers to the entire $i$-th row of the matrix.
Sets are represented using calligraphic font (e.g., $\mathcal{X}$).
The bracket notations $\left[\cdot\,;\cdot\right]$ and $\left[\cdot,\cdot\right]$ refer to vertical and horizontal concatenations, respectively.
The symbol $\otimes$ represent the Kronecker product.
The vector $\vec{e}^K_k$ represents the canonical basis vector of dimension $K$, where the $j$-th element $\vec{e}^K_k[j]$ is 1 if $j=k$ and 0 otherwise.
$\ell_{2,1}$ norm for $\left[\vec{a}, \vec{b}\right]$ with $\vec{a},\vec{b}\in\mathbb{R}^N$ is defined as $\|\left[\vec{a}, \vec{b}\right] \|_{2,1} \coloneqq \sum_{k=0}^{N-1} \sqrt{a[k]^2+b[k]^2}$.


\subsection{PACT imaging model}\label{back:operator}
In PACT, a short laser pulse illuminates the object, and a portion of the absorbed optical energy is converted into an IP distribution through photoacoustic effect.
The induced IP distribution, denoted as $p_0(\vr)\in L^2(\mathbb{R}^d)$, generates a pressure wavefield $p(\vr,t)\in L^2(\mathbb{R}^d \times [ 0,T))$ within the object, where $\vr\in\mathbb{R}^d$ and $t\in [0,T)$ represent the spatial and the temporal coordinates, respectively, and $T$ denotes the acquisition duration.
Here, $L^2$ denotes the space of square-integrable functions.
The photoacoustic wave propagation in an unbounded and lossless medium can be described by the following coupled first-order partial differential equations \cite{tabei2002k, treeby2010modeling}:
\begin{align} \label{eq:waveeqs}
\rho(\vr)\frac{\partial \bm{u}}{\partial t} (\vr, t) &= -\nabla p(\vr,t), \nonumber\\
\frac{1}{\rho(\vr)c^2(\vr)}\frac{\partial p}{\partial t} (\vr, t) &= -\nabla \cdot \bm{u}(\vr,t), \nonumber\\ p(\vr,0) &= p_0(\vr), \,\,\bm{u}(\vr,0) = \mathbf{0}, \mathrm{\,\,for\,\,} \bm{r}\in\mathbb{R}^d,\,\, t\in[0,T),
\end{align}
where $\bm{u}(\vr,t)\in \left[L^2(\mathbb{R}^d \times [ 0,T))\right]^d$, $c(\vr)\in L^2(\mathbb{R}^d)$, and $\rho(\vr)\in L^2(\mathbb{R}^d)$ are the acoustic particle velocity, SOS distribution, and ambient density distribution, respectively. 
In this study, the ambient density distribution is assumed to be known and constant.
This  is justified because   density variations can be  generally considered negligible in soft biological tissue.

A set of ultrasound transducers positioned around the object measures the  pressure wavefield that propagates out of the irradiated object.
The pressure time-trace recorded at the $n$-th receiving transducer will be denoted as $g_n(t) \in L^2([0,T))$. 
This quantity is related to the photoacoustic wavefield as $g_n(t) = m_np(\cdot,t) \coloneqq \int_{\mathbb{R}^2} \chi_n(\vr') p(\vr', t) d\vr'$,
where $\chi_n(\vr')$ is the indicator (or characteristic) function of the support of the $n$-th transducer. 
Here, $m_n:L^2(\mathbb{R}^2 \times [0,T) )\to L^2([0,T))$ 
represents the corresponding spatial sampling operator at the $n$-th transducer.

\subsection{Discrete forward modeling}
Although wave propagation occurs in 3D, for simplicity, 2D wave physics is considered in this study.
Let $\vr_k = (x_i, y_j) \in \mathbb{R}^{2}$ denote a point on a 2D Cartesian grid, where $k = i \cdot N_y + j$, with $i = 0, \dots, N_x - 1$ and $j = 0, \dots, N_y - 1$. 
Here, $N_x$ and $N_y$ represent the number of discrete points along the $x$ and $y$ directions, respectively. 
The finite-dimensional representations of the functions $p_0(\vr)$ and $c(\vr)$, denoted as $\vp\in\mathbb{R}^N$ and $\vc\in\mathbb{R}^N$, can be expressed as $\vp[ k ] \coloneqq p_0(\vr_k)$ and $\vc[ k ] \coloneqq c(\vr_k)$, respectively.
Let $t_l \in\mathbb{R}^+$  for $l = 0, 1, \dots, N_t-1$ specify the temporal sampling points, with $N_t$ representing the number of time-sampling points. 
The discrete pressure vector $\vd \in \mathbb{R}^{N_r\cdot N_t}$, representing samples from all receiving transducers, is defined as
$\vd[k] \coloneqq g_n(t_l)$ for $k = N_r \cdot l + n$, where $l = 0,\dots,N_t-1$ and $n=0,\dots,N_r-1$. 
Here, $N_r$ represents the total number of the receiving transducers.
Considering these discrete sampling effects, 
a discrete-to-discrete (D-D) forward model for PACT can be expressed as  \cite{huang2013full, matthews2017joint}:
\begin{equation}\label{eq:DD}
    \vd = \M\Hc_{\vc}\vp.
\end{equation}
Here, $\Hc_{\vc}\in \mathbb{R}^{N\cdot N_t \times N}$ represents a discretized wave-propagation operator, which explicitly depends on $\vc$, and $\M\in\mathbb{R}^{N_r\cdot N_t \times N\cdot N_t}$ serves as a discrete sampling operator that evaluates the discretized pressure field at the transducer locations.
The associated inverse problem considered is to estimate $\vp$ and $\vc$, given the the sampled measurement data $\vd$.

\section{Constrained JR for PACT} \label{method}
\subsection{Review of penalized JR problem}
For the JR problem in PACT, an optimization framework can be established to estimate $\vp$ and $\vc$ simultaneously based on the D-D measurement model in Eq. \eqref{eq:DD}.
In general, this JR problem can be expressed as \cite{huang2016joint}:
\begin{equation}\label{eq:JR}
    \vp^*, \vc^* = \argmin_{\vp, \vc} J(\vp, \vc) +\lambda_{\vp} R_{\vp}(\vp) + \lambda_{\vc} R_{\vc}(\vc),
\end{equation}
where $J: \mathbb{R}^N \times \mathbb{R}^N \to \mathbb{R}$ is a data fidelity term that can be defined as:
\begin{equation*}
J(\vp,\vc) = \frac{1}{2}\| \M \Hc_{\vc} \vp - \vd\|_2^2,
\end{equation*}
and the penalty terms $R_{\vp}(\vp)$ and $R_{\vc}(\vc)$ are possibly non-smooth functions. 
The scalars and $\lambda_{\vp}$ and $\lambda_{\vc}$ are the corresponding penalty parameters. 
Eq.\ \eqref{eq:JR} can be solved using alternating optimization methods \cite{huang2016joint, attouch2010proximal}, where the minimization is performed alternately over $\vp$ and $\vc$ while holding the other fixed at each step. 

\subsection{Constrained JR problem formulation}
Rather than using penalty terms, prior knowledge about the object can be directly incorporated by constraining the estimates of $\vp$ and $\vc$ to belong to admissible sets that reflect the object's known properties.
This study focuses on the following three canonical object-based constraints:
\begin{itemize}
    \item \textbf{Support constraint \cite{vaswani2010modified, wang2010sparse}.} This constraint limits the reconstruction region to the spatial extent of the object, excluding the water bath where IP and SOS values are known with high accuracy. This reduces the number of parameters to estimate for the IP and SOS distributions, while preserving image features inside the support
    \item \textbf{Bound constraint \cite{chen2012non, sidky2012convex}.} This constraint restricts the values of the  IP and SOS estimates to physically plausible ranges.
    \item \textbf{TV constraint \cite{esser2018total,zhang2016investigation}.} This constraint imposes a bound on the total variation (TV) norm of the IP and SOS estimates, limiting the overall variation while preserving important structural details, such as boundaries and edges.
\end{itemize}
For simplicity, let $\vf$ represent either $\vp$ or $\vc$. 
The admissible sets associated with the support, bound, and TV constraints for $\vf$ are denoted as $\X^S_{\vf}$, $\X^B_{\vf}$, and $\X^{TV}_{\vf}$, respectively.
The overall admissible set $\X_{\vf}$, which combines these three types of constraints, is given by $\X_{\vf} = \X^{S}_{\vf} \cap \X^{B}_{\vf} \cap \X^{TV}_{\vf}$.
The constrained optimization problem for JR can then be expressed as:
\begin{equation}\label{eq:JR2}
    \vp^*, \vc^* = \argmin_{\vp, \vc}J(\vp, \vc) \,\, \mathrm{\,\,subject\,\, to \,\,} \vp \in \X_{\vp} \,\, \mathrm{\,\,and \,\,} \vc \in \X_{\vc}.
\end{equation}
The numerical studies described in Section \ref{sec:ns} seek to assess the extent to which these constraints can  stabilize the joint estimation of the IP and SOS distributions.

In following subsections, the specific formulations of $\X^S_{\vf}$, $\X^B_{\vf}$, and $\X^{TV}_{\vf}$ are provided.




\subsubsection{Support constraint}
Let $\Omega \subset \mathbb{R}^2$ denote the set of locations within the  imaged object.
The element $\vf[k]$ for $\vr_k\notin \Omega$ is assumed to be a known constant, denoted as $v_b$, throughout the iterative image reconstruction process. 
The associated admissible set  $\X^S_{\vf}$ can be defined as:
\begin{equation}\label{eq:supp}
    \X^S_{\vf} = \{\vf\mid \vf[k] = v_{b} \mathrm{\,\,for \,\, all \,\,} \vr_k\notin \Omega\}.
\end{equation}

\subsubsection{Bound constraint}
Let $l(\vr)$ and $u(\vr)$ denote the physically plausible lower and upper bound functions for $v(\vr)$, respectively, such that $l(\vr)\le v(\vr) \le u(\vr)$.
Their finite-dimensional representations are denoted as $\vec{l}$ and $\vec{u}$, where $\vec{l}[k]\coloneqq l(\vr_k)$ and $\vec{u}[k]\coloneqq u(\vr_k)$ for $k=0,1,\ldots,N-1$.
The associated admissible set $\X^B_{\vf}$ can be defined as:
\begin{equation}
    \X^B_{\vf} = \{\vf\mid \vec{l}[k] \le \vf[k] \le \vec{u}[k] \mathrm{\,\,for\,\,} k= 0,1, \dots, N-1\}.
\end{equation}

\subsubsection{TV constraint} \label{back:tv}
Let $\Dx\in\mathbb{R}^{N\times N}$ and $\Dy\in\mathbb{R}^{N\times N}$ denote the 2D forward finite difference operators, approximating variations along the $x$-direction (vertical axis) and the $y$-direction (horizontal axis), respectively.
Specifically, $\Dx$ and $\Dy$ are defined as $\Dx=\vec{D}_{1d}^x \otimes \vec{I}_{N_y}$ and $\Dy=\vec{I}_{N_x} \otimes \vec{D}_{1d}^y$ where $\vec{I}_{N_x}\in\mathbb{R}^{N_x\times N_x}$ and $\vec{I}_{N_y}\in\mathbb{R}^{N_y\times N_y}$ are identity matrices and $\vec{D}_{1d}^x\in\mathbb{R}^{N_x\times N_x}$, $\vec{D}_{1d}^y\in\mathbb{R}^{N_y\times N_y}$ are 1D forward finite difference operators defined those entries are given by
\begin{equation*}
\vec{D}_{1d}^x[i_x, j_x] = \begin{cases}
-1 & \text{if } j_x = i_x,\, 0 \leq i_x \leq N_x-2\\
1 & \text{if } j_x = i_x+1,\, 0 \leq i_x \leq N_x-2\\
0 & \text{otherwise},
\end{cases}
\end{equation*}
and 
\begin{equation*}
\vec{D}_{1d}^y[i_y, j_y] = \begin{cases}
-1 & \text{if } j_y = i_y,\, 0 \leq i_y \leq N_y-2\\
1 & \text{if } j_y = i_y+1,\, 0 \leq i_y \leq N_y-2\\
0 & \text{otherwise}.
\end{cases}
\end{equation*}


Finally, the finite difference matrices $\Dx$ and $\Dy$ are scaled by a scalar weight $w_{\vf}$ to balance the TV norms of $\vp$ and $\vc$, such that the weighted operators are $w_{\vf}\Dx$ and $w_{\vf}\Dy$.
Finally, the weighted TV norm of $\vf$ is defined as $\| \vf \|_{wTV} = \| \left[
    w_{\vf}\Dx\vf, w_{\vf}\Dy\vf\right]\|_{2,1}$.
The associated admissible set $\X^{TV}_{\vf}$ is then defined as
\begin{equation}
    \X^{TV}_{\vf} = \{\vf \mid \vf\in\mathbb{R}^{N}, \| \vf\|_{wTV} \le \tau_{\vf}  \},
\end{equation}
where  $\tau_{\vf}\in\mathbb{R}^+$ is a threshold value for the weighted TV norm.

\subsection{Image reconstruction}\label{method:recon}
The constrained JR problem in Eq.\ \eqref{eq:JR}, considering the support, bound, and TV constraints, can be expressed as the following unconstrained problem:
\begin{equation}\label{eq:JR3}
    \vp^*, \vc^* = \argmin_{\vp,\vc} J(\vp,\vc) + I_{\X_{\vp}}(\vp) + I_{\X_{\vc}}(\vc),
\end{equation}
Here, the indicator function $I_{\mathcal{V}}$ for an arbitrary set $\mathcal{V}$ is defined as:
\begin{equation}
    I_{\mathcal{V}}(\vf) =
    \left\{
    \begin{array}{ll}
        0, & \mathrm{\,\,if \,\,} \vf \in \mathcal{V} \\
        \infty, & \mathrm{\,\,otherwise}.
    \end{array}
    \right.
\end{equation}
With $\x \coloneqq \left[\vp;\vc\right]$, the objective function $J(\vp,\vc)$ can be equivalently denoted as $J(\x)$, with the understanding that this overloaded notation implies  $J(\x)=J(\vp,\vc)$.
Additionally, a block matrix $\D\in\mathbb{R}^{4N\times 2N}$ is defined as:
\begin{equation}
    \D = \left[ 
    \begin{array}{cc}
        \D_{\vp} & \vec{0}_{2N} \\
        \vec{0}_{2N} & \D_{\vc} \\
    \end{array} 
    \right], \mathrm{ \,\,where \,\,}
    \D_{\vp} = \left[
    \begin{array}{c}
        \Dx_{\vp} \\  \Dy_{\vp}
    \end{array}
    \right], \mathrm{\,\, and \,\,}
    \D_{\vc} = \left[
    \begin{array}{c}
        \Dx_{\vc} \\  \Dy_{\vc}
    \end{array}
    \right].
\end{equation}
Here, $\vec{0}_{2N}\in\mathbb{R}^{2N\times2N}$ is the zero matrix.
Introducing an auxiliary variable $\z\coloneqq\left[\zpx;\zpy;\zcx;\zcy\right]\in\mathbb{R}^{4N}$, where $\zpx,\zpy,\zcx,\zcy\in\mathbb{R}^N$, the JR problem in Eq.\ \eqref{eq:JR2} can be re-expressed by incorporating the linearly coupled equality constraint $\D\x=\z$ as:
\begin{equation} \label{eq:prim1}
    \x^* = \argmin_{\x} J(\x) + I_{\X^{SB}}(\x) + I_{\Z}(\z) \mathrm{\,\,such\,\, that\,\, } \D\x = \z,
\end{equation}
where the set $\X^{SB}$ is defined as
\begin{equation}
    \X^{SB} = \{\x = \left[\vp;\vc \right]  \mid \vp \in \X^{S}_{\vp} \cap \X^{B}_{\vp}, \vc \in\X^{S}_{\vc}\cap \X^{B}_{\vc}\},\nonumber
\end{equation}
and the set $\Z$ is defined as
\begin{equation}
    \Z = \{\z = \left[\zpx;\zpy;\zcx; \zcy\right] \mid \|\left[\zpx, \zpy\right]\|_{2,1}\le \tau_{\vp},  \|\left[\zcx, \zcy\right]\|_{2,1}\le \tau_{\vc}\}. \nonumber
\end{equation}
This approach is inspired by Ren \textit{et al.} \cite{ren2024simultaneous}, in which a constrained optimization problem was formulated for the joint reconstruction of activity and attenuation in time-of-flight positron emission tomography.

The primal problem in Eq.\ \eqref{eq:prim1} can be solved by the augmented Lagrangian method \cite{nocedal1999numerical, bertsekas2014constrained}.
The augmented Lagrangian is obtained by adding a quadratic penalty to the Lagrangian associated with the primal problem in Eq.\ \eqref{eq:prim1} as 
\begin{equation}\label{eq:AL1}
    \mathcal{L}_{\rho}\left(\x, \z, \vl\right) = J(\x) + I_{\X}(\x) +I_{\Z}(\z) + \vl^T \left(\D\x-\z\right) + \frac{\rho}{2}\|\D\x-\z\|^2_2,
\end{equation}
where $\vl \in \mathbb{R}^{4N}$ is the dual variable (Lagrangian multiplier) and $\rho\in\mathbb{R}^+$ is a penalty parameter.
For any $\rho \geq 0$, the associated saddle point problem is formulated as
\begin{equation}\label{eq:saddle}
  \x^*,\z^*,\vl^* =  \argmin_{\x,\z} \max_{\vl}\mathcal{L}_{\rho}\left(\x, \z, \vl\right),
\end{equation}
where $\x^*$ is a minimizer of the  primal problem in Eq.\ \eqref{eq:prim1}.
The alternating direction method of multipliers (ADMM) \cite{boyd2011distributed, parikh2014proximal} approximates a saddle point of Eq.\ \eqref{eq:saddle} by alternating a minimization step with respect to $\x$ and $\z$ with an ascent update with respect to $\vl$.
Specifically, the $k$-th ADMM iteration is
\begin{align}
\label{eq:admm:primal_update}
\x^{k+1} &= \argmin_{\x} \mathcal{L}_{\rho}\left(\x, \z^k, \vl^{k}\right), \\
\z^{k+1} &= \argmin_{\z} \mathcal{L}_{\rho}\left(\x^{k+1}, \z, \vl^{k}\right), \\
\vl^{k+1} &= \vl^{k} + \rho\left(\D\x^{k+1} - \z^{k+1}\right).
\end{align}
Algorithmic details of the ADMM employed in this study are provided in \ref{App}.

\section{Numerical studies}
\label{sec:ns}
Four  computer-simulation studies were conducted to systematically investigate the effectiveness of object constraints in stabilizing joint estimation of the IP and SOS distributions. 
These studies involved anatomically realistic numerical breast phantoms (NBPs) and a physics-based method for simulating the associated PACT measurement data. 
The studies addressed the following:
\begin{itemize}
    \item \textbf{Study I}: A strict inverse crime study where the discretization parameters for both the measurement simulation and image reconstruction are identical, with no measurement noise included.
    \item \textbf{Study II}: A study involving data inconsistencies associated with different discretization parameters for measurement simulation and image reconstruction, along with the inclusion of measurement noise.
    \item \textbf{Study III}: An extension of Study II that explores various breast types, each with a different level of discrepancy between the initial estimates of SOS and the true SOS distributions.
    \item \textbf{Study IV}: An extension of \textbf{Study II} that incorporates heterogeneous acoustic attenuation (AA) during measurement simulation, while ignoring AA or assuming a simplified homogeneous AA model during image reconstruction.
\end{itemize}

\subsection{Generation of phantoms} \label{ns:phantom}
Two-dimensional computer simulation studies were conducted using 2D distributions of the IP and SOS, which characterized the to-be-imaged object.
These 2D distributions were extracted from a large ensemble of anatomically realistic 3D optoacoustic NBPs, generated following the framework established by Park \textit{et al.} \cite{park2023stochastic} for virtual imaging trials of PACT. 
These 3D NBPs, each representing one of four breast types based on the BI-RADS classifications \cite{radiology2013acr}—A) almost entirely fatty breasts, B) breasts with scattered fibroglandular density, C) heterogeneously dense breasts, and D) extremely dense breasts—reflect clinically relevant variability in anatomy, shape, and optical and acoustic tissue properties.
The IP distribution within an NBP was generated via a Monte Carlo simulation of photon transport using the \texttt{MCX} software \cite{fang2009monte, yu2018scalable}.
The illumination geometry was chosen to produce a nearly uniform fluence distribution on the breast surface. It consisted of 20 arcs uniformly distributed over an hemispherical surface, each emitting 1 mJ laser pulse at a wavelength of 800 nm. Additional details on the optical simulation can be found in Section 4.2 of Park \textit{et al.} \cite{park2023stochastic}.

Two-dimensional cross-sectional (coronal plane) distributions of IP, SOS, and AA were extracted from a 3D NBP with a voxel size of 0.125 mm at an elevation within the central region of the breast.
However, in 2D PACT imagers that employ elevation-focused transducers, there exists a slice thickness over which structures are averaged. 
To emulate this effect, the values of IP, SOS, and AA were vertically averaged within a thin slab of thickness 1.375 mm, rather than being taken from a single plane.
For regions outside the breast, the extracted 2D IP distribution values were set to 0 kPa, assuming negligible optical absorption in the water bath at the 800 nm laser wavelength. 
Similarly, the extracted 2D SOS and AA distribution values in the water bath were assigned constant values of 1.5206 mm/$\mu$s and 0.00022 dB/MHz$^y$mm, respectively, where $y$ is the constant exponent for AA.
It is important to note that the extracted 2D AA distributions were only used in \textbf{Study IV}. 
The resolution of these extracted 2D distributions was maintained at 0.125 mm.

In summary, a total of 360 3D NBPs, 90 for each breast type, were used in this study. 
Two distinct sets were established, each consisting of 2D distributions of IP, SOS, and AA that were extracted from this ensemble of NBPs.
One set, referred to as the \textit{calibration} set, was used to determine the bounds and TV weight values (as detailed in Section \ref{ns:constraint}), compute the ensemble-averaged SOS value (as described in Section \ref{ns:study3}), and calculate the ensemble-averaged AA value for D-type breasts (as detailed in Section \ref{ns:study4}).
The other set, referred to as the \textit{test} set, was used to evaluate reconstruction performance in \textbf{Studies I---IV.}
These sets were defined as follow:
\begin{itemize}
    \item The \textit{calibration} set consists of 2D distributions of IP, SOS, and AA extracted from 280 NBPs, with 70 NBPs from each breast type. Four distinct 2D slices were extracted from each NBP at different elevations, spaced at least 3.25 mm apart to ensure structural variability among the slices from the same NBP. In total, this results in 1,120 2D slices. The extracted 2D distributions, originally at a pixel size of 0.125 mm, were down-sampled via bilinear interpolation to a 512-by-512 grid with a pixel size of 0.32 mm.
    \item The \textit{test} set consists of 2D distributions of IP, SOS, and AA (for \textbf{Study IV}) extracted from 80 NBPs, with 20 NBPs from each breast type. 
    One 2D slice was extracted from each NBP, resulting in a total of 80 2D slices.
    The extracted slices were down-sampled using bilinear interpolation to either a 512-by-512 grid with a pixel size of 0.32 mm for \textbf{Study I} (applied only to D-type breasts), or a 1024-by-1024 grid with a pixel size of 0.16 mm for \textbf{Studies II---IV} (used for all breast types).
\end{itemize}
Throughout the study, the 512-by-512 grid is referred to as the \textit{coarse} grid, and the 1024-by-1024 grid as the \textit{fine} grid.

\subsection{Simulation of pressure data} \label{ns:measurement}
A 2D stylized acoustic detection system was considered to simulate pressure wavefield based on the extracted 2D distributions of IP and SOS (including AA for \textbf{Study IV}). 
The acoustic detection system consisted of a circular transducer array of radius 72 mm, with 512 idealized point-like receiving transducer elements evenly spaced along the array. 
The acquisition duration was 112 $\mu$s, which is long enough to capture all the transmitted waves originated within the field of view.

Two different 2D acoustic wave propagation models were employed to simulate measurement data: a lossless model that neglected AA effects for \textbf{Studies I---III}, and a \textbf{lossy} model that accounted for heterogeneous AA distributions for \textbf{Study IV}.
The lossless acoustic wave equation, represented as coupled first-order partial differential equations in Eq.\ \eqref{eq:waveeqs}, was used for the first three studies. 
For the last study, the lossy acoustic wave equation, incorporating power-law absorption modeling, was employed \cite{treeby2010modeling, huang2013full}. 

Both wave propagation models were discretized using the time-domain pseudospectral \textit{k}-space method \cite{treeby2010photoacoustic}.
The Courant-Friedrichs-Lewy number was set to 0.318 for maximum SOS of 1.59 mm/$\mu$s, resulting in a timestep size of 0.064 $\mu$s for the \textit{coarse} grid (\textbf{Study I}) and 0.032 $\mu$s for the \textit{fine} grid (\textbf{Studies II---IV}).
To minimize boundary reflections caused by the finite size of the computational domain, perfectly matched layers with a thickness of 6.4 mm were applied at each edge of the grid \cite{berenger1994perfectly}.
The Python package, \texttt{j-Wave} (version 0.2.0) \cite{stanziola2023j}, was used for the numerical solution of the lossless acoustic wave propagation equation. 
Since \texttt{j-Wave} does not natively support the lossy case, specific modifications were made to incorporate power-law absorption effects for the lossy acoustic wave equation.

\subsection{Image reconstruction}\label{ns:recon}

Image reconstruction was performed on the \textit{coarse} grid across all numerical studies.
The pressure wavefields estimated during image reconstruction were simulated over 1750 timesteps of size 0.064 $\mu$s. 
For \textbf{Studies II---IV}, the measurement data, originally acquired over 3500 timesteps, were sub-sampled to match the 1750 timesteps.

The IP and SOS estimates for image reconstruction were initialized as homogeneous distributions, with all elements of $\vp$ set to 0 kPa and all elements of $\vc$ were set to 1.5206 mm/$\mu$s, representing the water bath values. 
These homogeneous distributions are denoted as $\vf^w$, where $\vf^w$ refers to $\vp^w$ for the IP and $\vc^w$ for the SOS distribution.

In the case in which TV constraints were applied, image reconstruction was performed by use of the ADMM method presented in Section \ref{method:recon}. 
When TV constraints were not used, the same algorithm employed to compute the primal update in ADMM (minimization with respect to the IP and SOS in Eq.\ \eqref{eq:admm:primal_update}) was used for image reconstruction. 
More details can be found in \ref{App}.


\subsection{Object constraints} \label{ns:constraint}
Here, the employed support, bound, and TV constraints for the IP and SOS are described.
The terms \textit{tight} and \textit{loose} in the following sections denote the tightness level of the admissible sets associated with these constraints.
A \textit{tight} constraint is based on precise prior knowledge of the object’s properties, ensuring that the JR solutions exactly match the true characteristics of the object. 
In contrast, a \textit{loose} constraint provides a broader solution space to account for uncertainties in prior information.

\subsubsection{Support constraint} \label{ns:support}
The \textit{tight} support constraint was used exclusively in \textbf{Study I}.
This constraint defined the support region to exactly match true object's breast region, assuming perfect knowledge of the object's boundary.
However, in practice, object boundary segmentation \cite{jin2022signal, mandal2016visual} can only provide an estimate of the true boundary.
To account for this uncertainty, the \textit{loose} support constraint was implemented for \textbf{Studies II---IV}, where the support region was conservatively expanded by dilating the true object boundary by 3.2 mm.
In all cases, the admissible sets associated with the support constraint for the IP and SOS distributions were identical for each object (i.e., $\mathcal{X}^{S}_{\vp} = \mathcal{X}^{S}_{\vc}$)


\subsubsection{Bound constraint} \label{ns:bound}
The \textit{tight} bound constraint was used exclusively in \textbf{Study I}. 
For both the IP and SOS distributions, the lower and upper bounds were set to the minimum and maximum values observed in the true distributions of each subject, respectively.
The \textit{loose} bound constraint was employed in \textbf{Studies II---IV}.  
A sufficiently large upper bound of $1\times10^{16}$ kPa was applied, which is within the limits of single precision. 
The lower bound was set to 0 kPa, as IP values are physically constrained to be non-negative.
For the SOS distribution, the \textit{loose} bound constraint was designed based on the minimum and maximum values observed across an ensemble of SOS distributions in the \textit{calibration} set.
The ensemble minimum and maximum SOS values, serving as lower and upper bounds, were 1.41 mm/$\mu$s and 1.59 mm/$\mu$s, respectively.

\subsubsection{TV constraint}\label{ns:tv}
The TV weight values $\weip$ for IP and $\weic$ for SOS were determined using the \textit{calibration} set. 
First, the ensemble-averaged unweighted TV norms were computed for each distribution, defined as $\| \vf \|_{TV} = \| \left[\Dx\vf, \Dy\vf\right]\|_{2,1}$, where $\vf$ represents either $\vp$ (IP) or $\vc$ (SOS). The resulting values were 0.91 for $\vp$ and 590.08 for $\vc$. The weights, $\weip$ and $\weic$, were then set to the reciprocals of these values---1/0.91 and 1/590.08, respectively.

The threshold values for the weighted TV norms, $\tau_{\vp}$ for IP and $\tau_{\vc}$ for SOS, were determined using the true object's weighted TV norm values, $t_{\vp}$ for IP and $t_{\vc}$ for SOS.
These thresholds were defined as:
$\tau_{\vp} = \alpha t_{\vp}$ and
$\tau_{\vc} = \alpha t_{\vc}$,
where $\alpha$ is a positive real number that controls the tightness of the TV constraints for both IP and SOS.
The value of $\alpha$ was adjusted for each study, with specific values selected for \textbf{Studies I---IV} based on the study objectives.

\subsection{Evaluation metrics} \label{eval}
The performance of the constrained JR methods was evaluated using the normalized root mean squared error (NRMSE). 
For a given reconstructed property map $\vf$ (either $\vp$ or $\vc$), the NRMSE can be defined as:
\begin{equation}
    \mathrm{NRMSE} = \frac{\|\vf^{t} - \vf\|_2}{\|\vf^{t}-\vf^w\|_2},
\end{equation}
where $\vf^{t}$ represents the true object distribution, and $\vf^w$, as previously defined, represents the initial homogeneous distribution corresponding to the water bath values.
Additionally, the NRMSE was calculated only within the breast region of interest. 
Let $\Omega^B\subset\mathbb{R}^2$ denote the set of a 2D Cartesian grid points within the breast.
The NRMSE within the targeted breast region (NRMSEb) can be given by: 
\begin{equation}
\mathrm{NRMSEb} = \frac{\sqrt{\sum_{k: \vr_k \in \Omega^B}(\vf^t[k] - \vf[k])^2}}{\sqrt{\sum_{k: \vr_k \in \Omega^B}(\vf^t[k] - \vf^w[k])^2}}.
\end{equation}
The mean and standard deviation (SD) of both the NRMSE and NRMSEb were computed on the \textit{test} set.
For \textbf{Studies II---IV}, the NRMSE and NRMSEb were computed after down-sampling the true distributions to the \textit{coarse} grid, where image reconstructions were performed.

\begin{figure}[tbhp]
  \centering
      \includegraphics[width=0.97\textwidth]{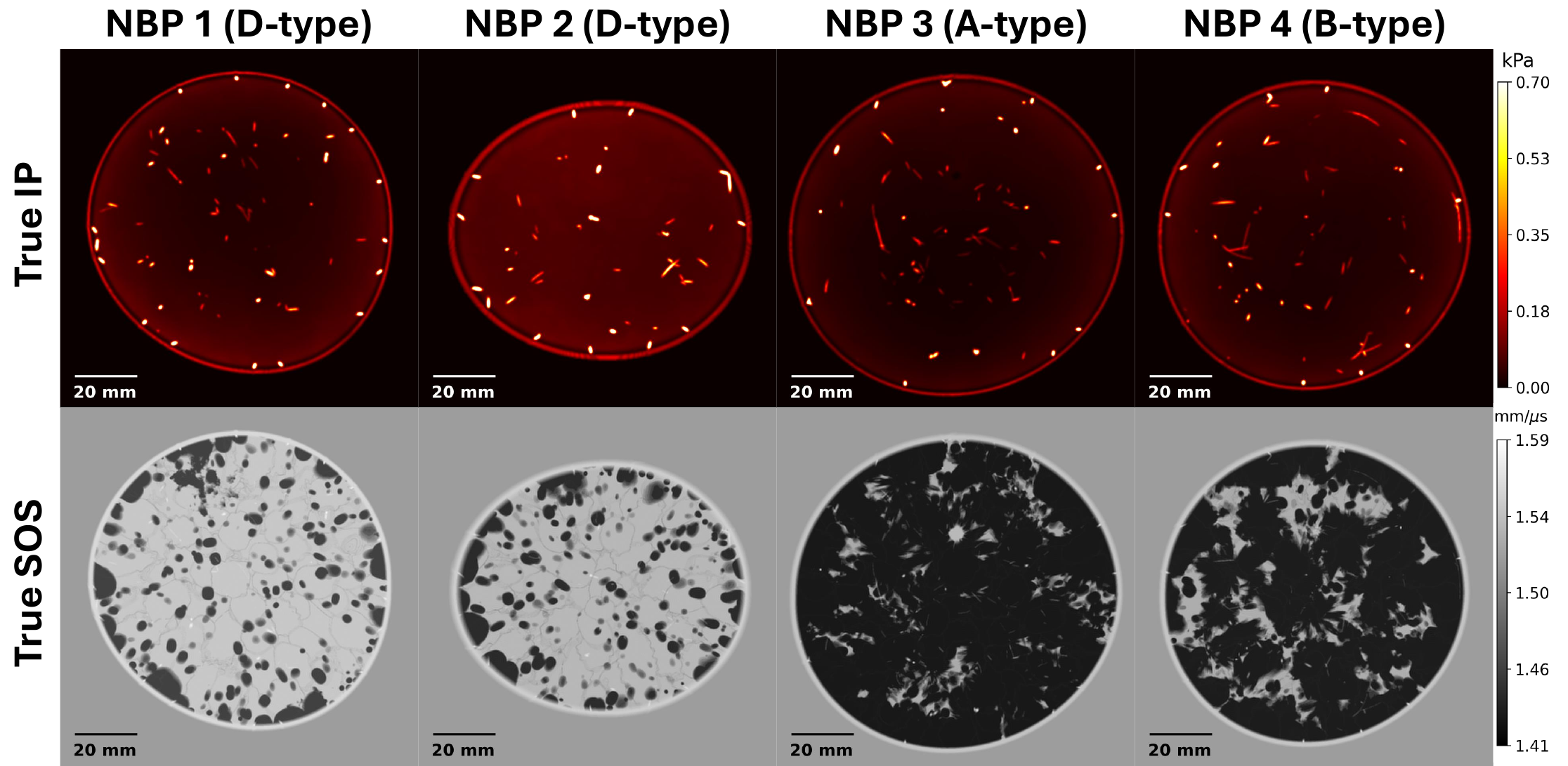} 
  \caption{Examples of the numerical breast phantoms employed in the numerical studies.  The top row shows the true IP distributions for four different breast types, labeled NBP 1 to NBP 4. The bottom row shows the corresponding true SOS distributions. The IP colorbar was set with a minimum of 0 kPa and a maximum based on the 99.9th percentile of pixel values from an ensemble of IP images. For the SOS colorbar, the bounds were set to 1.41 mm/$\mu$s and 1.59 mm/$\mu$s, which are the ensemble’s minimum and maximum values. These colorbars are consistently used in all figures in Section \ref{results}. All images, including those in subsequent figures, were cropped from 163.84mm-by-163.84mm to a window size of 119.04mm-by-119.04mm.}
  \label{fig:true}
\end{figure}

\subsection{Study designs}

\subsubsection{\textbf{Study I}: Impact of canonical object constraint under inverse crime conditions} \label{ns:study1}
The effectiveness of canonical object constraints was investigated systematically under ideal conditions.
To isolate the impact of these constraints and establish a benchmark for optimal performance, a strict \textit{inverse crime} approach was adopted, which ensures data consistency between measurement simulation and reconstruction by maintaining identical wave propagation models, computational grids, and temporal sampling rates, as detailed in Sections \ref{ns:measurement} and \ref{ns:recon}. Additionally, no measurement noise was considered.

Image reconstruction was performed across eight different scenarios, varying by the presence or absence of the \textit{tight} support, bound, and TV constraints.
For scenarios involving the \textit{tight} TV constraint, $\alpha$ was set to 1. 
Note that in these cases, the TV ball radii for $\vp$ and $\vc$ are equal to their respective TV norms of the true object.

The ensemble-averaged the NRMSE and NRMSEb were evaluated using 20 slices of D-type breasts in the \textit{test} set. 
NBP 1 and NBP 2 in Fig.\ \ref{fig:true} serve as examples of the 2D distributions of true IP and SOS used in this study.

\subsubsection{\textbf{Study II}: Impact of data inconsistency} \label{ns:study2}
The effectiveness of canonical object constraints was evaluated under data inconsistency conditions, introduced by the different discretization parameters between the measurement simulation and reconstruction, as detailed in Sections \ref{ns:measurement} and \ref{ns:recon}. 
Additionally, independent and identically distributed (i.i.d) Gaussian noise with zero mean was added to the simulated measurement data, with signal-to-noise ratio (SNR) set at 20, 15, and 10 dB.

Image reconstructions were performed using a combination of the TV, \textit{loose} support, and \textit{loose} bound constraints.
For the TV constraint, four tightness levels were considered, with $\alpha$ set to 0.8, 1.0, 1.2, and 1.5.

The ensemble-averaged NRMSE and NRMSEb were assessed using 20 slices of D-type breasts in the \textit{test} set across the three SNR levels and four different $\alpha$ values.
NBP 1 and NBP 2 in Fig.\ \ref{fig:true} serve as examples of the true IP and SOS distributions used in this study.

\begin{figure}[tbhp]
  \centering
      \includegraphics[width=0.97\textwidth]{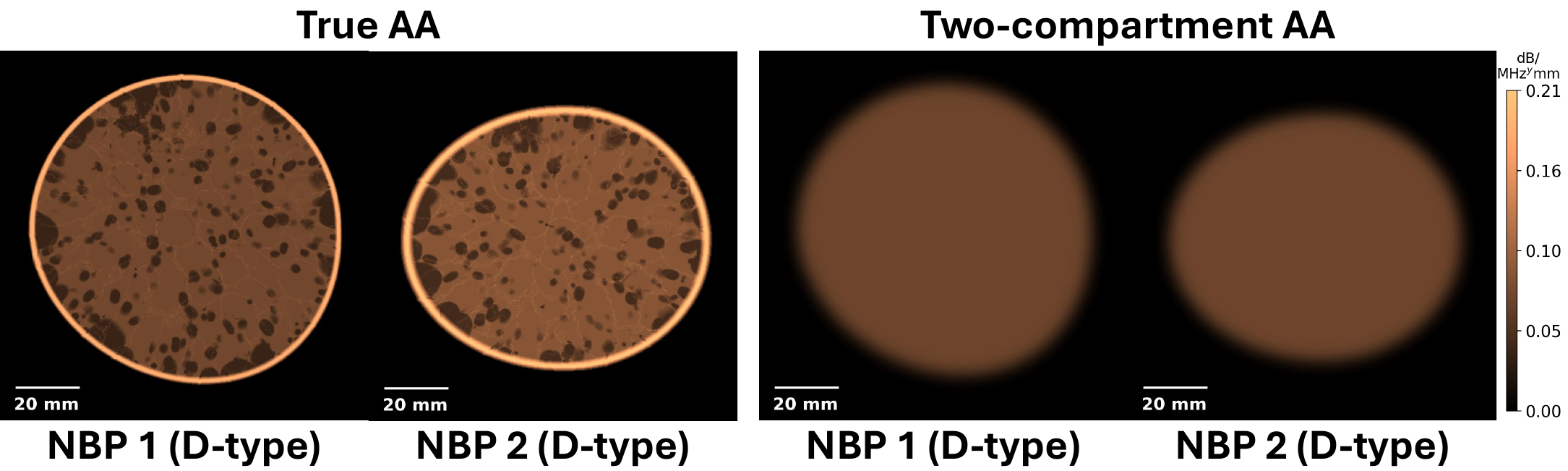} 
  \caption{Example of AA distributions used in \textbf{Study IV}. The left column shows the true AA distributions of NBP 1 and NBP 2 corresponding to the IP and SOS distributions presented in Fig.\ \ref{fig:true}, which were considered when simulating measurement data.
   The right column shows the two-region AA distributions that were assumed when performing JR to estimate the IP and SOS distributions.}
  \label{fig:trueaa}
\end{figure}

\subsubsection{\textbf{Study III}: Impact of tissue composition and initial SOS distribution}\label{ns:study3}
In \textbf{Studies I} and \textbf{II}, only D-type breasts were employed.
The ensemble-averaged SOS value within the breast region for these cases, evaluated from the \textit{calibration} set, was 1.524 mm/$\mu$s, which is close to the initial constant SOS estimate of 1.5206 mm/$\mu$s (i.e., $\vc^w$ in Section \ref{ns:recon}).
However, significant SOS variations exist across breast types due to differences in fatty and fibroglandular tissue composition. 
The ensemble-averaged within-breast SOS values evaluated using the \textit{calibration} set were 1.456 mm/$\mu$s for A-type , 1.471 mm/$\mu$s for B-type, and 1.499 mm/$\mu$s for C-type breasts.
These values indicate a larger discrepancy between the initial SOS estimate and the true SOS distributions for A, B, and C-type breasts than D-type breasts.  
This discrepancy suggests that using a homogeneous initial SOS of 1.5206 mm/$\mu$s could pose challenges for the non-convex JR problem in breast types other than D-type. 

To investigate the effectiveness of the use of canonical object constraints for A-, B-, and C-type breasts, data inconsistency conditions due to the discretization parameter mismatches (as conducted in \textbf{Study II}) were considered.
Additionally, i.i.d. Gaussian noise was added to measurements, with SNR set at 15 dB.
Image reconstructions were performed using a combination of the \textit{loose} support, \textit{loose} bound, and TV  constraints, with $\alpha$ of 0.8 and 1.2.

The ensemble-averaged NRMSE and NRMSEb values were assessed using 60 2D slices, with 20 slices each from the A, B, and C breast types in the \textit{test} set.
NBP 3 and NBP 4 in Fig.\ \ref{fig:true} serve as examples of the 2D distributions of true IP and SOS used in this study.

\subsubsection{\textbf{Study IV}: Impact of AA variations}\label{ns:study4}
In \textbf{Studies I--III}, measurement data were simulated assuming a lossless medium, ignoring ultrasound absorption, which can be a significant physical factor in soft, vascularized tissues \cite{parker2022power}.
To address this limitation, power law absorption effects were incorporated into the measurement data simulation, as detailed in \ref{ns:measurement}. 

Two approaches were designed to investigate the effectiveness of the use of canonical object constraints in the presence of significant modeling errors due to heterogeneous AA distributions.
In the first approach, AA was neglected during reconstruction, meaning that while heterogeneous AA distributions were considered in simulation of the measurement data, the forward model used to estimate the pressure field during reconstruction did not account for AA.
In the second approach, a fixed two-region AA distribution was employed to partially account for AA effects within the breast. 
This model assigned constant values to the breast region and the water bath. 
These values were kept fixed and not subject to estimation during the reconstruction process.
The two-region AA distribution was constructed as follows: the breast region, defined as 3.2 mm narrower than the true object's boundary, was assigned a single AA value of 0.0734 dB/MHz$^y$mm---the ensemble-averaged AA value from 2D slices of D-type breasts within the breast region in the \textit{calibration} set.
The exterior was assigned the water's AA value of 0.00022 dB/MHz$^y$mm. 
This model was then smoothed using a 2.56mm-by-2.56mm Gaussian kernel. 

Data inconsistency conditions stemming from mismatches in the discretization parameters  were considered, as conducted in \textbf{Studies II} and \textbf{III}.
Additionally, i.i.d. Gaussian noise was added to measurements, with SNR set at 15 dB.
Image reconstructions were performed using a combination of the \textit{loose} support, \textit{loose} bound, and TV  constraints, with $\alpha$ of 0.8 and 1.2.
The exponent power of AA ($y$ in Section \ref{ns:phantom}) was set at 1.3635 both in the measurement simulation and in the second approach where two-region AA distributions were considered.

The ensemble-averaged NRMSE and NRMSEb values were assessed using 20 slices of D-type breasts in the \textit{test} set.
NBP 1 and NBP 2 in Fig.\ \ref{fig:true} serve as examples of the 2D distributions of true IP and SOS used in this study.
The corresponding true AA distributions are shown in Fig.\ \ref{fig:trueaa}, along with the two-region AA distributions.



\begin{figure}[tbhp]
  \centering
  \includegraphics[width=0.97\textwidth]{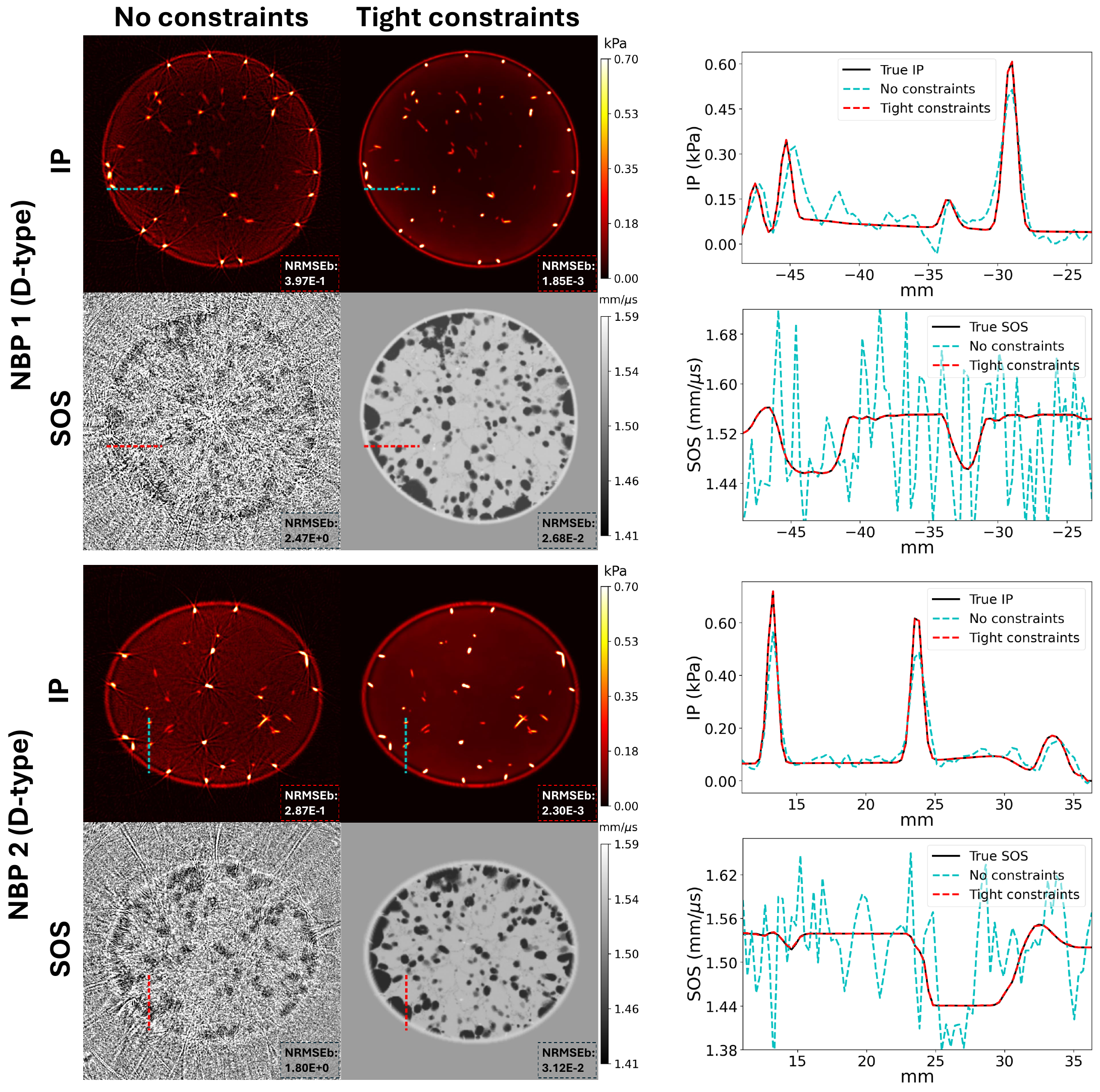}
  \caption{Reconstructed IP and SOS distributions for \textbf{Study I}, a strict inverse crime study using the same discretization parameters for both measurement simulation and image reconstruction, and no measurement noise included. The first row corresponds to NBP 1 (D-type), and the second row corresponds to NBP 2 (D-type). In each subpanel, the left and right columns show the reconstructed IP (top) and SOS (bottom) distributions corresponding to the cases without constraints and with \textit{tight} constraints (including support, bound, and TV), respectively. Line profiles along locations indicated by dashed lines on the IP and SOS images compare the reconstructed and true distributions. Unconstrained reconstructions show severe distortions in both IP and SOS, whereas tightly constrained reconstructions achieve high accuracy.}
  \label{fig:study1}
\end{figure}

\section{Results} \label{results}
\subsection{\textbf{Study I}: Impact of canonical object constraint under inverse crime conditions} 
The impact of object constraints on the stability and accuracy of the JR was investigated, with data consistency maintained between the measurement simulation and reconstruction process.
Specifically, data generation and image reconstruction were performed on the \textit{coarse} grid (timestep size of 0.064 $\mu$s). Measurement data did not include noise.

Examples of reconstructed IP and SOS distributions, along with the corresponding line profiles, are shown in Fig.\ \ref{fig:study1}, comparing two cases: one without constraints and another with \textit{tight} constraints, including support, bound, and TV.
As expected, the reconstructed IP and SOS distributions were severely distorted in the case where no constraints were employed, as shown in the first column of Fig.\ \ref{fig:study1}.
Notably, peak intensities in the IP distribution appeared significantly out of focus, and overall variations in the reconstructed SOS did not effectively capture those of the true SOS distribution, as shown in the line profiles in Fig.\ \ref{fig:study1}.
In contrast, applying \textit{tight} support, bound, and TV constraints ($\alpha$ = 1) resulted in highly accurate estimates of both the IP and SOS distributions, as demonstrated in the second column of Fig.\ \ref{fig:study1}.
The IP and SOS line profiles shown in the right panel of Fig.\ \ref{fig:study1} display a close alignment between the true and reconstructed values.
These examples demonstrate that \textit{tight} support, bound, and TV constraints can effectively define a solution space for IP and SOS, sufficiently mitigating the non-uniqueness of the JR problem.

\begin{table}[t]
\footnotesize
\caption{\textbf{Study I}: The ensemble-averaged NRMSE and NRMSEb values for the reconstructed IP and SOS distributions in the presence or absence of \textit{tight} support, bound, and TV constraints.}\label{tab:study1_metric}
\begin{center}
  \begin{tabular}{|c|c|c|c|c|c|c|c|c|c|} \hline 
    \multicolumn{2}{|c|}{\multirow{3}{*}{}} & \multicolumn{4}{c|}{No TV} & \multicolumn{4}{c|}{\textit{Tight} TV ($\alpha$=1)}  \\\cline{3-10}
      \multicolumn{2}{|c|}{}& \multicolumn{2}{c|}{NRMSE} & \multicolumn{2}{c|}{NRMSEb} &  \multicolumn{2}{c|}{NRMSE} & \multicolumn{2}{c|}{NRMSEb} \\ \cline{3-10}
      \multicolumn{2}{|c|}{}& Mean & SD &  Mean & SD & Mean & SD &Mean & SD  \\ \hline
    No Support $\&$ & IP & 50.1 &11.9& 39.3 &9.04 & 27.8& 15.5& 27.8 &15.5 \\
    No Bound & SOS & 432 &138 & 265& 80.7 & 260 &95.5 &50.3& 36.4\\\hline     
    \textit{Tight} Support $\&$ & IP & 35.1 &13.1& 35.1 &13.1& 0.222 &0.073 & 0.222 &0.073 \\
    No Bound& SOS &436 &103& 435& 103& 2.72& 0.591 & 2.72& 0.591 \\\hline
    No Support $\&$&IP& 23.7 &3.52& 22.2& 3.01 & 8.02& 5.58 & 8.01 &5.58 \\
    \textit{Tight} Bound & SOS& 238& 34.8& 108 &9.61 & 74.4 &17.1 & 11.7 &5.11 \\\hline
    \textit{Tight} Support $\&$ &IP&3.47 &0.676& 3.47& 0.676 & 0.207& 0.051 & 0.207& 0.051  \\
    \textit{Tight} Bound & SOS& 45.2 &7.96& 45.2 &7.96 & 2.64& 0.452& 2.64& 0.452 \\\hline    
  \end{tabular}
  \smallskip 
\parbox{0.95\linewidth}{
    \small 
Note 1: The mean and SD values of NRMSE and NRMSEb are scaled by \(10^{2}\).
}
\end{center}
\end{table}

\begin{figure}[!htpb]
  \centering
    \includegraphics[width=0.97\textwidth]{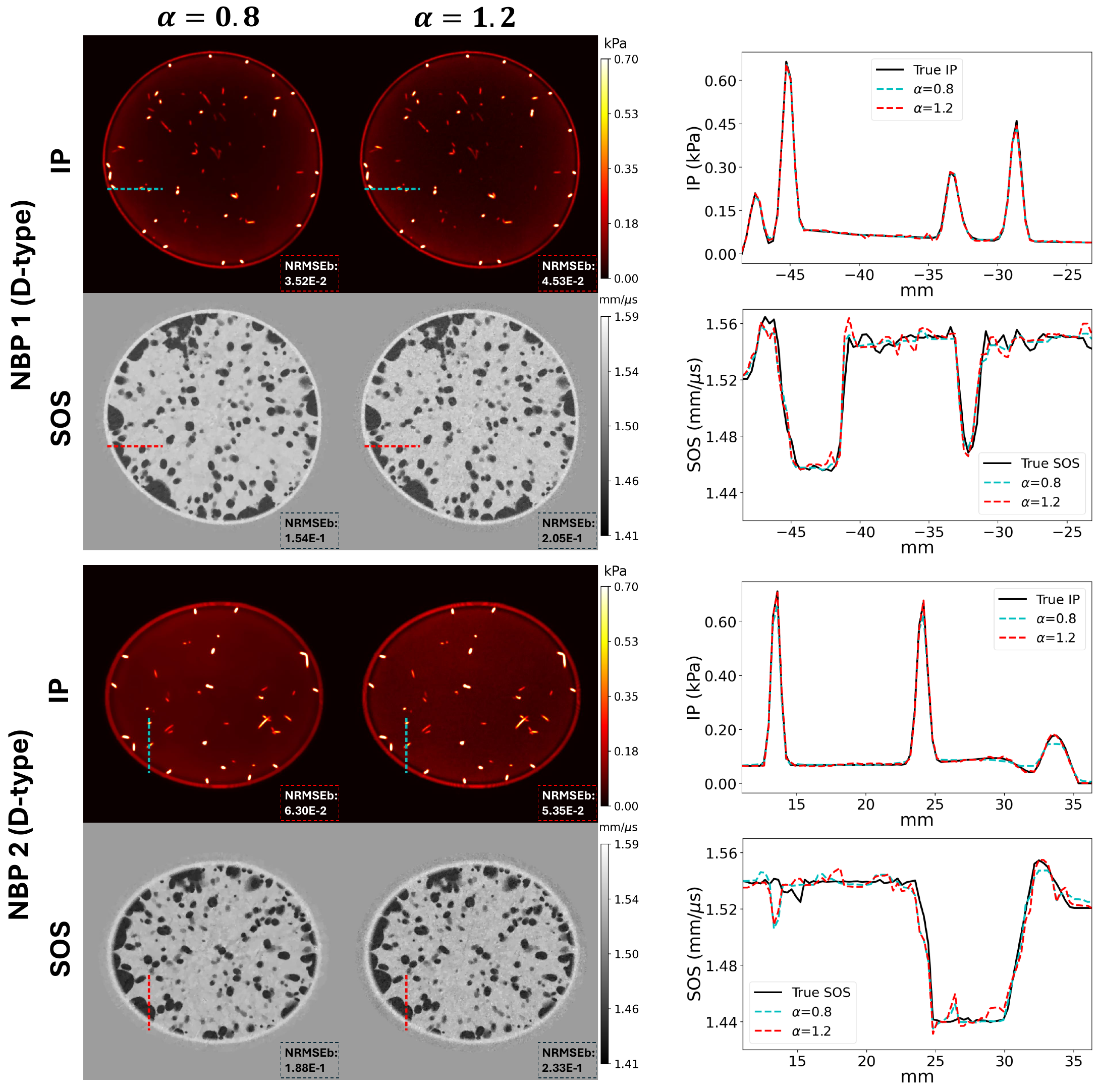}   
  \caption{Reconstructed IP and SOS distributions for \textbf{Study II}, which included modeling errors (stemming from the use of different discretization parameters for measurement simulation and image reconstruction) and measurement noise (SNR: 15 dB). The first row corresponds to NBP 1 (D-type), and the second row corresponds to NBP 2 (D-type). All reconstructions applied the \textit{loose} support and \textit{loose} bound constraints, while the TV constraint was applied with $\alpha$ of 0.8 and 1.2. In each subpanel, the left and right columns show the reconstructed IP (top) and SOS (bottom) distributions corresponding to $\alpha$ = 0.8 and $\alpha$ = 1.2, respectively. Line profiles along locations indicated by dashed lines on the IP and SOS images compare the reconstructed and true distributions. Results demonstrate object constraints' effectiveness in stabilizing the JR and achieving accurate solutions despite data inconsistency.}
  \label{fig:study2}
\end{figure}

Table \ref{tab:study1_metric} presents the ensemble-averaged NRMSE and NRMSEb values, computed on the \textit{test} set consisting 20 D-type breasts across eight scenarios, each representing different combinations of applying or not applying the three \textit{tight} constraints: support, bound, and TV.
As expected, applying more constraints led to greater improvements in the reconstruction accuracy of both IP and SOS distributions, as indicated by the NRMSE and NRMSEb values.
Notably, the effectiveness of the TV constraint was significantly enhanced when combined with support and bound constraints, leading to improved reconstruction accuracy.

Similar accuracy was observed between the case using all \textit{tight} constraints and the case using only \textit{tight} support and \textit{tight} TV constraints without the bound constraint, suggesting that the bound constraint may not be as critical when TV and support constraints are already \textit{tight}. 
However, in scenarios where the TV constraint was absent, the combination of support and bound constraints significantly outperformed the use of the support constraint alone, as indicated by the NRMSE and NRMSEb values.
These findings suggest that while the bound constraint may appear less important when the TV constraint is \textit{tight}, it becomes crucial when the TV constraint is \textit{loose}. 

\begin{table}[t]
\footnotesize
\caption{\textbf{Study II}: The ensemble-averaged NRMSE and NRMSEb values for the reconstructed IP and SOS distributions across various SNRs.}\label{tab:study2_metric}
\begin{center}
  \begin{tabular}{|c|c|c|c|c|c|c|c|c|c|} \hline 
   \multicolumn{2}{|c|}{\multirow{2}{*}{}} & \multicolumn{4}{c|}{IP} & \multicolumn{4}{c|}{SOS}   \\ \cline{3-10}  
\multicolumn{2}{|c|}{} & \multicolumn{2}{c|}{NRMSE} & \multicolumn{2}{c|}{NRMSEb} & \multicolumn{2}{c|}{NRMSE} & \multicolumn{2}{c|}{NRMSEb}\\ \hline
   SNR & $\alpha$ & Mean & SD & Mean & SD & Mean & SD & Mean & SD \\ \hline
    $\infty$ &1.0 & 1.34 &0.312& 1.23 &0.317& 8.88 &0.875  & 8.59&0.715 \\
   \hline
    \multirow{4}{*}{20} &0.8 & 5.98 &2.75& 5.89 &2.74& 16.2 &4.59& 15.8 &4.37 \\
     &1.0 & 2.66&0.379& 2.55&0.383& 13.2 &1.00& 12.8 &0.842 \\
     &1.2 & 3.73 &0.524& 3.62&0.532& 16.5&1.24& 15.9 &1.08 \\
     &1.5 & 4.88 &0.534& 4.77&0.540& 20.9 &1.49& 19.9 &1.37\\    \hline    
    \multirow{4}{*}{15} &0.8 & 5.92 &2.23& 5.79&2.22& 17.7&3.14& 17.3 &2.92\\
     &1.0 & 3.99 &0.497& 3.86 &0.503& 17.2 &1.46& 16.7 &1.40\\
     &1.2 & 5.24 &0.529& 5.09 &0.540& 21.5&1.78& 20.7 &1.63 \\
     &1.5 & 6.86 &0.738& 6.70 &0.750& 26.3 &2.04& 25.2 &1.92 \\    \hline        
    \multirow{4}{*}{10} &0.8 & 6.89 &1.56& 6.73 &1.57& 22.3 &1.94& 21.7 &1.77 \\
     &1.0 & 6.21 &0.786& 6.05 &0.786& 23.9 &2.63& 23.2 &2.58\\
     &1.2 & 7.48 &0.583& 7.28 &0.584& 27.7 &2.12& 26.7 &2.10 \\
     &1.5 & 9.32 &0.754& 9.07 &0.760& 34.2 &2.77& 32.8 &2.53 \\    \hline    
  \end{tabular}
  \smallskip 
\parbox{0.95\linewidth}{
    \small 
Note: Mean and SD values are scaled by \(10^{2}\).
}
\end{center}
\end{table}

\subsection{\textbf{Study II}: Impact of data inconsistency}

The effectiveness of using object constraints was investigated in the presence of discretization error as well as measurement noise. Data generation was performed on the \textit{fine} grid (timestep size of 0.032 $\mu$s), while the \textit{coarse} grid (timestep size of 0.064 $\mu$s) was used for image reconstruction. Measurement data were corrupted with i.i.d. Gaussian noise at SNR of 20, 15, and 10 dB.
The \textit{loose} support and \textit{loose} bound constraints were applied, with the TV constraintusing $\alpha$ of 0.8, 1.0, 1.2, and 1.5.

Examples of reconstructed IP and SOS distributions, along with the corresponding line profiles, are shown in Fig.\ \ref{fig:study2} for $\alpha$ of 0.8 and 1.2, using data with an SNR of 15 dB. 
The reconstructed IP and SOS distributions closely approximated their corresponding true distributions.
The reconstructed IP appear slightly over-smoothed at peak intensities, especially for NBP 2, as shown in the line profile in Fig.\ \ref{fig:study2} when the TV constraint is overly \textit{tight} ($\alpha$=0.8).

\begin{figure}[!tbhp]
  \centering
    \includegraphics[width=0.97\textwidth]{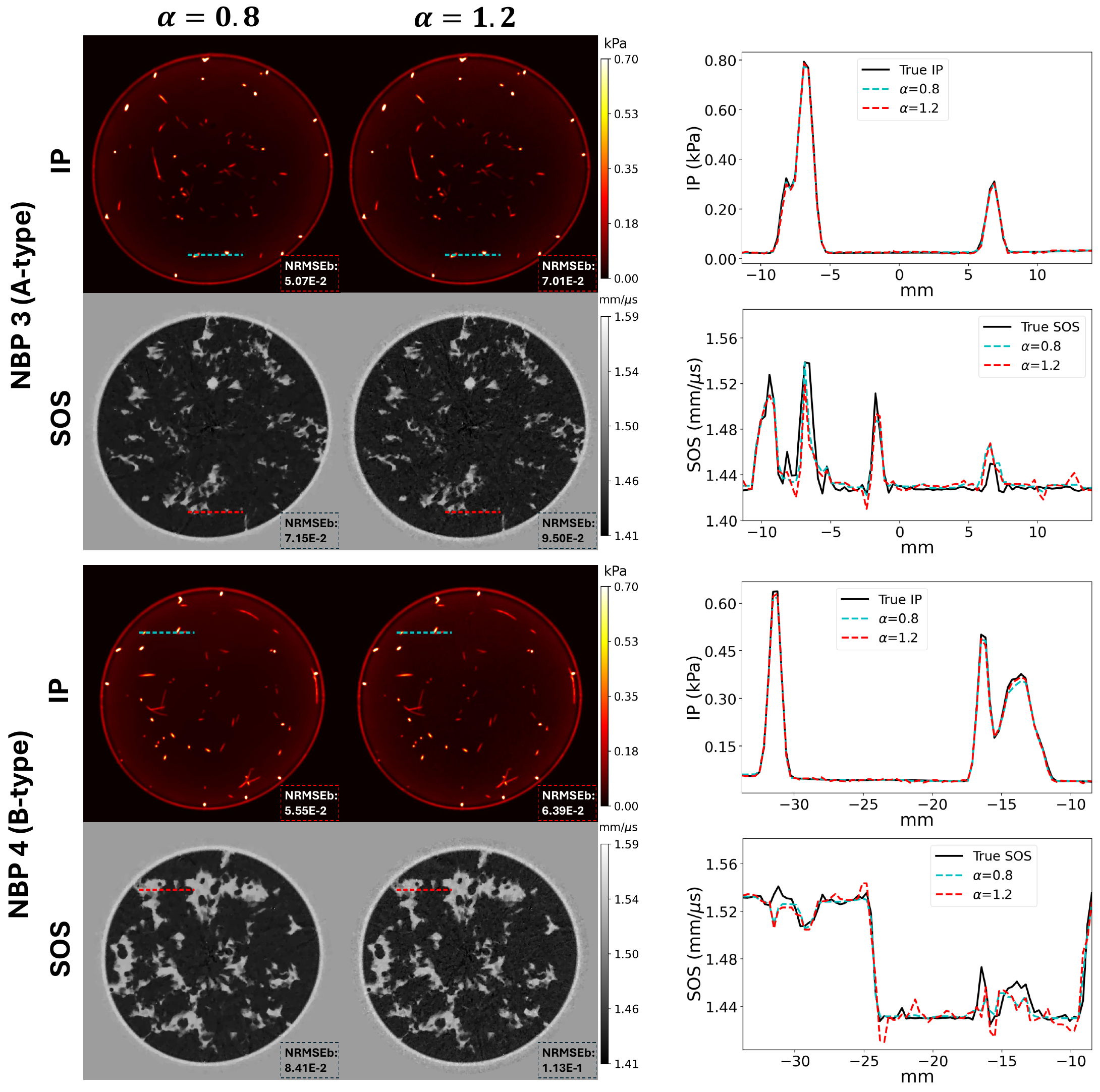}   
\caption{Reconstructed IP and SOS distributions for \textbf{Study III}, which included modeling errors (stemming from the use of different discretization parameters for measurement simulation and image reconstruction) and measurement noise (SNR: 15 dB).
The first row corresponds to NBP 3 (A-type), and the second row corresponds to NBP 4 (B-type). All reconstructions applied \textit{loose} support and \textit{loose} bound constraints, while the TV constraint was applied with $\alpha$ of 0.8 and 1.2. In each subpanel, the left and right columns show the reconstructed IP (top) and SOS (bottom) distributions corresponding to the cases with $\alpha$ = 0.8 and $\alpha$ = 1.2, respectively. Line profiles along locations indicated by dashed lines on the IP and SOS images compare the reconstructed and true distributions. Despite initializing with a uniform SOS of 1.5206 mm/$\mu$s—significantly deviating from the true SOS distributions of A-type and B-type NBPs—the constrained JR method produced IP and SOS distributions that closely approximated the true values.}
\label{fig:study3}
\end{figure}

Table \ref{tab:study2_metric} presents the ensemble-averaged NRMSE and NRMSEb values, computed on the \textit{test} set consisting 20 D-type breasts across SNRs of 10, 15, and 20 dB, for $\alpha$ values of 0.8, 1.0, 1.2, 1.5.
These results are compared against a noise-free case, with discretization parameter mismatch.
For this noise-free case, the \textit{loose} support, \textit{loose} bound, and \textit{tight} TV constraint ($\alpha$ = 1) were applied.
Although reconstruction accuracy, as indicated by the NRMSE and NRMSEb values, decreased with each 5 dB reduction in SNR, the rate of degradation remained moderate across the SNR levels.
Moreover, while different $\alpha$ values led to variations in reconstruction accuracy at the same SNR, these differences were relatively small.
These findings suggest that the canonical object constraints can effectively stabilize the JR across various levels of Gaussian noise, even without exact prior knowledge of the true object's support, bound, and TV norm values, provided a reasonable range of constraint values is used.

\begin{table}[h]
\footnotesize
\caption{\textbf{Study III}: The ensemble-averaged NRMSE and NRMSEb values for the reconstructed IP and SOS distributions stratified by breast type (A, B, and C), with measurement noise at an SNR of 15 dB. For type D breasts, these metrics are provided in Table \ref{tab:study2_metric}.}\label{tab:study3_metric}
\begin{center}
  \begin{tabular}{|c|c|c|c|c|c|c|c|c|c|} \hline 
   \multicolumn{2}{|c|}{\multirow{2}{*}{}} & \multicolumn{4}{c|}{IP} & \multicolumn{4}{c|}{SOS}   \\ \cline{3-10}  
\multicolumn{2}{|c|}{} & \multicolumn{2}{c|}{NRMSE} & \multicolumn{2}{c|}{NRMSEb} & \multicolumn{2}{c|}{NRMSE} & \multicolumn{2}{c|}{NRMSEb}\\ \hline
   Type & $\alpha$ & Mean & SD & Mean & SD & Mean & SD & Mean & SD \\ \hline
    \multirow{2}{*}{A} &0.8 & 6.04 &1.26& 5.91 &1.24 & 8.51 &1.88 & 8.41 &1.88 \\
      &1.2 & 6.83 &0.594& 6.69 &0.606  & 10.7 &2.24  & 10.5 &2.17 \\     \hline        
    \multirow{2}{*}{B} &0.8 & 5.69 &1.12 & 5.57 &1.09 & 11.1 &2.93 & 11.0 &2.91 \\
     &1.2 &  7.09 &0.483 & 6.95& 0.487 & 14.3& 3.42 & 14.0& 3.37 \\
     \hline        
    \multirow{2}{*}{C}  &0.8 &  5.37 &1.10 & 5.26 &1.10 & 13.4 &1.66 &  13.3 &1.65 \\    
      &1.2 & 6.98 &0.523 & 6.86& 0.532 & 18.5 &1.97 & 18.0 &1.95 \\
     \hline           
  \end{tabular}
  \smallskip 
\parbox{0.95\linewidth}{
    \small 
Note: Mean and SD values are scaled by \(10^{2}\).
}
\end{center}
\end{table}

\subsection{\textbf{Study III}: Impact of tissue composition and initial SOS distribution}
The effectiveness of using object constraints was investigated across breast types (A to D) in the presence of discretization error as well as measurement noise.
Data generation was performed on the \textit{fine} grid (timestep size of 0.032 $\mu$s), while the \textit{coarse} grid (timestep size of 0.064 $\mu$s) was used for image reconstruction.
Measurement data were corrupted with i.i.d. Gaussian noise (SNR: 15 dB).
The \textit{loose} support, \textit{loose} bound, and TV constraints, with $\alpha$ of 0.8 and 1.2, were applied.
The discrepancy between initial SOS estimates and true object values progressively increases from D to A. 

Examples of reconstructed IP and SOS distributions for representative A and B-type breasts (NBP 3 and NBP 4), along with the corresponding line profiles, are shown in Fig.\ \ref{fig:study3}. 
The reconstructed IP distributions closely matched the true maps in terms of overall structure and appearance.
The reconstructed SOS distributions effectively captured the structural details and heterogeneity of the true distributions.
Table \ref{tab:study3_metric} presents the ensemble-averaged NRMSE and NRMSEb values, computed on the \textit{test} set consisting of 20 breasts each of types A, B, and C. 
Values for D-type are provided in Table \ref{tab:study2_metric}.
Notably, no significant discrepancies in NRMSE and NRMSEb were observed across the different breast types.
These findings suggest that despite variations in tissue composition across different breast types, the support, bound, and TV constraints were still effective in stabilizing the JR process without requiring highly accurate initial estimates of the SOS distributions.
However, it should be noted that more accurate initial guesses could potentially further enhance reconstruction accuracy.

\begin{figure}[p]
  \centering
    \includegraphics[width=0.97\textwidth]{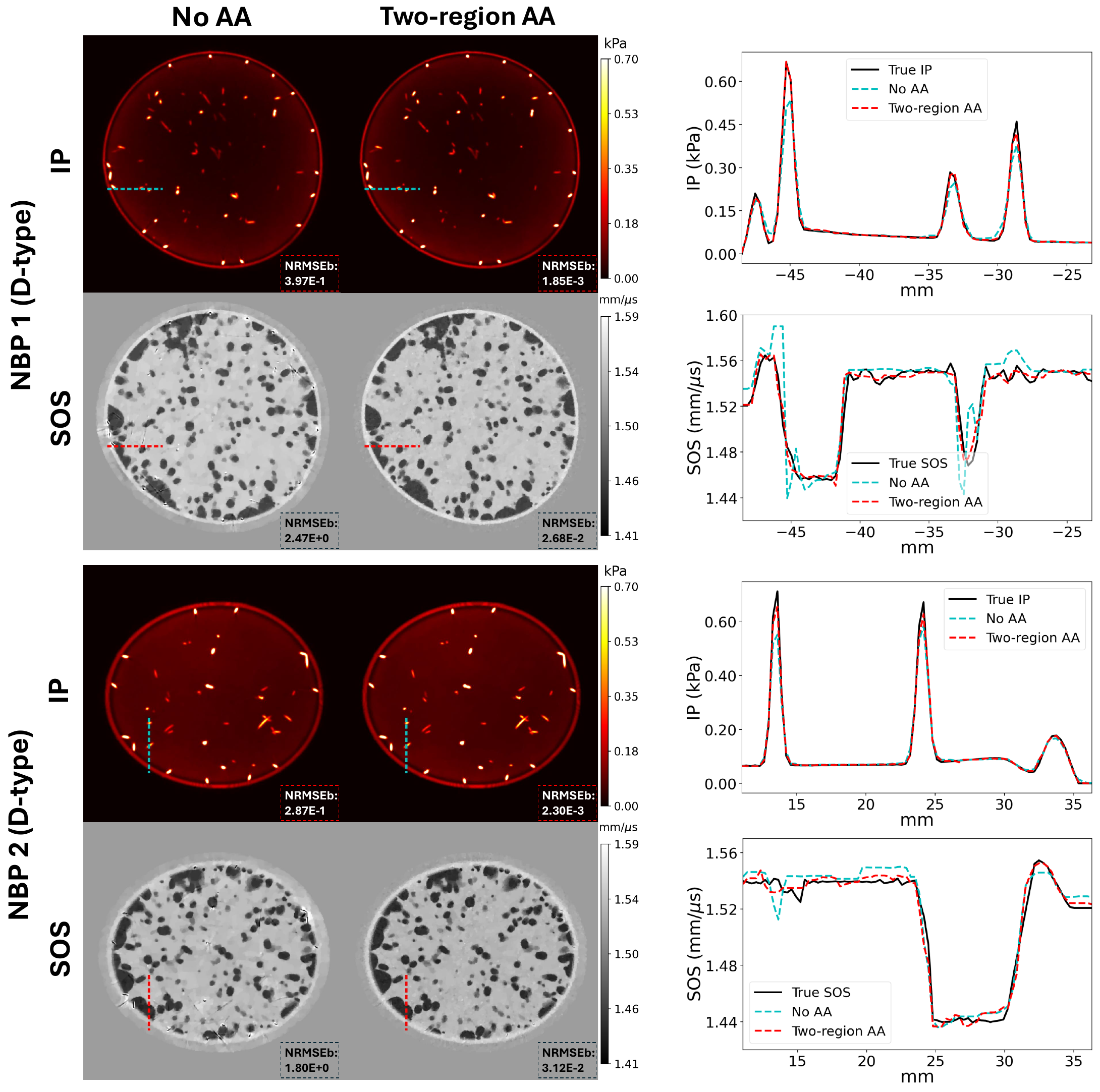}   
    \caption{Reconstructed IP and SOS distributions for \textbf{Study IV}, which included modeling errors (stemming from the use of different discretization parameters for measurement simulation and image reconstruction) and measurement noise (SNR: 15 dB). Heterogeneous AA (Fig.\ \ref{fig:trueaa}) was considered to simulate measurement data. The first row corresponds to NBP 1 (D-type), and the second row corresponds to NBP 2 (D-type). All reconstructions applied \textit{loose} support and \textit{loose} bound constraints, and the TV constraint was applied with $\alpha$ of 0.8. In each subpanel, the left and right columns show the reconstructed IP (top) and SOS (bottom) distributions corresponding to the cases where AA distributions were ignored and where two-region models (Fig.\ \ref{fig:trueaa}) were used during reconstruction, respectively. Line profiles along locations indicated by dashed lines on the IP and SOS images compare the reconstructed and true distributions. Ignoring AA in the JR process resulted in underestimations near vessels and streak-like artifacts in the SOS distribution. However, these artifacts were mitigated when the two-region AA models were used along with object constraints. }
    \label{fig:study4}
\end{figure}


\subsection{\textbf{Study IV}: Impact of AA variations}
The effectiveness of using object constraints was investigated in the presence of modeling and discretization errors as well as measurement noise.
Measurement data were simulated under the assumption of a power-law acoustic attenuation model and  heterogeneous AA distributions; IP and SOS distributions were jointly estimated under the assumption of a pure acoustic medium (no AA) or the use of a  simplified to two-region AA models.
Additionally, data generation was performed on the \textit{fine} grid (timestep size of 0.032 $\mu$s), while the \textit{coarse} grid (timestep size of 0.064 $\mu$s) was used for image reconstruction.
Measurement data were corrupted with i.i.d. Gaussian noise (SNR: 15 dB).
The \textit{loose} support and \textit{loose} bound constraints were applied, while TV constraints were applied with $\alpha$ of 0.8 and 1.2.

Examples of the reconstructed IP and SOS distributions, along with the corresponding line profiles, are shown in Fig.\ \ref{fig:study4} for $\alpha$ = 0.8.
When the AA effect was neglected during reconstruction, the peak intensities of the IP distributions were underestimated and  distortions were present in the SOS distributions around vessels and skin layers.
In contrast, applying a two-region model to compensate for heterogeneous AA effects effectively reduced artifacts around vessels in the SOS distribution and mitigated the underestimation of IP.
The ensemble-averaged NRMSE and NRMSEb values in Table \ref{tab:study4_metric} for both $\alpha$ values of 0.8 and 1.2 further support these findings. 
This demonstrates that the use of object constraints, even with a simplified two-region AA model, remains effective in producing high-quality estimates.

However, it should be noted that the reconstruction accuracy, as measured by NRMSE and NRMSEb, was slightly lower than that evaluated in \textbf{Study II} (Table \ref{tab:study2_metric}), where AA effects were neglected in both measurement simulation and reconstruction. 
This suggests that the two-region AA model could not fully account for the heterogeneous AA effects of the true object—such as amplitude attenuation and phase velocity dispersion \cite{herzfeld2013absorption}.
For enhanced accuracy, future studies might explore joint reconstruction of IP and SOS along with AA to better address these complex absorption effects.


\begin{table}[!t]
\footnotesize
\caption{\textbf{Study IV}: The ensemble-averaged NRMSE and NRMSEb values for the reconstructed IP and SOS distributions, with measurement noise at an SNR of 15 dB.} \label{tab:study4_metric}
\begin{center}
  \begin{tabular}{|c|c|c|c|c|c|c|c|c|c|} \hline 
   \multicolumn{2}{|c|}{\multirow{2}{*}{}} & \multicolumn{4}{c|}{IP} & \multicolumn{4}{c|}{SOS}   \\ \cline{3-10}  
\multicolumn{2}{|c|}{} & \multicolumn{2}{c|}{NRMSE} & \multicolumn{2}{c|}{NRMSEb} & \multicolumn{2}{c|}{NRMSE} & \multicolumn{2}{c|}{NRMSEb}\\ \hline
  AA & $\alpha$ & Mean & SD & Mean & SD & Mean & SD & Mean & SD \\ \hline
    \multirow{2}{*}{None}& 0.8 & 14.4 & 2.04 & 14.3 & 2.03 &  31.5 &2.97 & 28.4 &1.98 \\
    & 1.2 & 14.4 & 2.07 & 14.3 &2.06 & 38.0 &3.29 & 35.4 & 2.84 \\\hline     
    \multirow{2}{*}{Two-region} & 0.8 & 6.55 &1.45 & 6.49 &1.45 & 19.5 & 2.85 & 18.9 & 2.38 \\
    & 1.2 & 6.41 &0.704 & 6.31 &0.713  & 23.6 &1.66 & 22.6 &1.41 \\\hline    
  \end{tabular}
  \smallskip 
\parbox{0.95\linewidth}{
    \small 
Note 1: Mean and SD are scaled by \(10^{2}\).

}
\end{center}
\end{table}





\section{Conclusions}
\label{sec:conclusions}
This work numerically investigated the JR problem in PACT using three types of canonical object constraints: support, bound, and TV.
Under strict inverse crime conditions, use of \textit{tight} support, bound, and TV constraints
resulted in  highly accurate reconstructions of both the IP and SOS distributions.
This appears to be the first reported study in which such accurate  estimates of IP and SOS were jointly reconstructed from PACT data alone with consideration of complex and anatomically realistic numerical phantoms. 
These results suggest that, for the considered class of objects, the object constraints could effectively mitigate the non-uniqueness of the JR problem.

The effectiveness of these canonical object constraints in stabilizing the JR problem was further investigated under various data inconsistency conditions.
These included different discretization parameters between measurement simulation and reconstruction, as well measurement noise.
Even with loosely defined support, bound and TV constraints,
accurate JR solutions were stably computed across a range of noise levels.
Additionally, when the SOS distributions for different breast types deviated significantly from the homogeneous initial estimate, the reconstructed IP and SOS distributions remained accurate.
Moreover, in the presence of modeling error associated with a heterogeneous AA distribution,
the object-constrained JR approach continued to produce high quality estimates.
These findings suggest that the use of canonical object constraints may facilitate the application of JR to practical problems in the field of PACT.

There exist several topics for future investigation. 
A systematic investigation of the proposed methods using a 3D measurement geometry with consideration of data inconsistencies associated with an unknown transducer response will be a natural follow-up study.
Additionally, adjunct ultrasound measurements \cite{matthews2017joint} or multiple acquisitions of PACT data corresponding to different illumination strategies can be incorporated into the JR framework.
The evaluation of object-constrained JR methods by use of experimental data represents another important topic of future research.

\section*{Data availability statement}
The numerical phantoms and associated simulated PACT measurement data employed in the reported studies will be made publicly available upon acceptance of the manuscript.

\section*{Acknowledgements}
This work was supported in part by National Institutes of Health grants EB034261 and EB031585.
M.A.A.\ thanks Professor Xiaochuan Pan for many insightful discussions regarding  constrained optimization formulations of image reconstruction that served to motivate this study.

\appendix
\section{Algorithm to solve the constrained JR problem} \label{App}
\subsection{JR algorithm with TV constraints} \label{App:ADMM}
The augmented Lagrangian in Eq.\ \eqref{eq:AL1} can be re-written by directly scaling $\vl$ by $\rho$, resulting in:
\begin{equation}\label{eq:AL2}
    \mathcal{L}_{\rho}\left(\x,\z, \vl\right) = J(\x) + I_{\X}(\x) +I_{\Z}(\z) + \frac{\rho}{2}\|\D\x-\z+\frac{\vl}{\rho}\|^2_2 -\frac{\rho}{2}\| \frac{\vl}{\rho}\|^2_2.
\end{equation}
The detailed steps of the ADMM iterations for the augmented Lagrangian in Eq.\ \eqref{eq:AL2} are explicitly described in Algorithm \ref{alg:admm}.
The iterative process is performed until the norms of primal and dual residuals, denoted as $p$ and $d$ respectively, are sufficiently small \cite{boyd2011distributed}.
The penalty parameter $\rho$ is adaptively tuned to balance between $p$ and $d$, ensuring that both residuals converge at a similar rate \cite{he2000alternating, wang2001decomposition, boyd2011distributed}.
\begin{algorithm}[!htp]
\caption{ADMM}
\label{alg:admm}
\begin{algorithmic}[1]
\STATE{Initialize $\rho\in\mathbb{R}^+$, $\x^0=\left[{\vp^0}^T,{\vc^0}^T\right]^T\in\X^{SB}$, $\z^0 \in \Z$, and $\vl^0=\D\x^0-\z^0$ }
\STATE{Set $\epsilon_{abs}\in\mathbb{R}^{+}$ , $\epsilon_{rel}\in\mathbb{R}^{+}$}
\FOR{$k\ge 0$}
\STATE{$\x^{k+1} = \argmin_{\x} J(\x) + I_{\X^{SB}} (\x) + \frac{\rho}{2} \|\D \x - \z^k + \frac{\vl^k}{\rho}\|_2^2$} \label{sub}
\STATE{$\z^{k+1} = \argmin_{\z} I_{\Z}(\z) + \frac{\rho}{2}\|\D \x^{k+1} - \z + \frac{\vl^{k}}{\rho}\|_2^2$} \label{subz}
\STATE{$\vl^{k+1} = \vl^{k} + \rho \left( \D\x^{k+1} -\z^{k+1}\right)$}
\STATE{$\epsilon^{pri} = \sqrt{4N}\epsilon^{abs} + \epsilon^{rel}\cdot\max\{\|\D\x^{k+1}\|_2, \|\z^{k+1}\|_2\} $}
\STATE{$\epsilon^{dual} = \sqrt{2N}\epsilon^{abs} + \epsilon^{rel}\|\D^T\vl^{k+1}\|_2 $}
\STATE{$p = \|\D\x^{k+1} - \z^{k+1}\|_2$, $d =\|\rho\D^T\left(\z^{k+1}-\z^{k}\right)\|_2 $}
\IF{$\left(p\le\epsilon^{pri} \mathrm{ \,\,and\,\, } d\le\epsilon^{dual}\right)$}
    \STATE{\textbf{return} $\x^{k+1}$}
\ENDIF
\IF{$p > 10 \cdot d$}
    \STATE $\rho \gets 2 \cdot \rho$
\ELSIF{$d > 10 \cdot p$}
    \STATE $\rho \gets \rho / 2$
\ENDIF
\ENDFOR
\end{algorithmic}
\end{algorithm}



\subsubsection{$\x$-minimization step.}
The nonconvex and nonsmooth sub-problem in Line \ref{sub} of Algorithm \ref{alg:admm} is solved using an alternating minimization method, as detailed in Algorithm \ref{alg:step 1}. 
Specifically, given $\z^k$ and $\vl^k$, two sub-problems corresponding to $\vp$ and $\vc$ are alternately solved. 
Each  of these sub-problems (Lines \ref{subsubp0} and \ref{subsubc} in Algorithm \ref{alg:step 1}) is solved iteratively, with further details provided in Algorithm \ref{alg:sub}.
The gradients of the data fidelity term $J$ with respect to $\vp$ and $\vc$ were computed through the reverse-mode automatic differentiation provided by the Python library \texttt{JAX} \cite{jax2018github}, which forms the basis of the \texttt{j-Wave} \cite{stanziola2023j}.
To accelerate convergence for these sub-problems,  Barzilai-Borwein (BB) method \cite{barzilai1988two} is employed to adjust the step-size adaptively, approximating the inverse Hessian based on changes in the gradient between iterations. 
Furthermore, an \textit{Armijo}-type nonmonotone line search is applied to refine the BB step size. 
This line search is specifically designed for the projected gradient methods, where the projection is performed onto a closed and convex set \cite{birgin2000nonmonotone}.
The projection set, denoted as $\X_{\vf}^{SB}$, is always convex and closed. Its definition depends on whether the support or bound constraints are applied:
\begin{itemize}
    \item $\X_{\vf}^{SB} = \X_{\vf}^{S} \subseteq\mathbb{R}^N$ if no bound constraint is applied,
    \item $\X_{\vf}^{SB} = \X_{\vf}^{B} \subseteq\mathbb{R}^N$ if no support constraint is  applied,
    \item $\X_{\vf}^{SB} = \mathbb{R}^N$ if neither support nor bound constraint is not applied.
\end{itemize}
Given machine precision, all projections are performed in $\mathbb{R}^N$, which is closed and convex.

Note that when the TV constraint is not applied, the algorithm simplifies to the alternating minimization scheme outlined in Algorithm \ref{alg:step 1}, where the parameter $\rho$ is set to 0.



\begin{algorithm}[!htp]
\caption{Alternating optimization-based algorithm to solve the sub-problem with respect to $\x$ in Algorithm \ref{alg:admm}}
\label{alg:step 1}
\begin{algorithmic}[1]
\STATE{Initialize $\vp^{(k,0)} \gets \vp^{k}$ and $\vc^{(k,0)} \gets \vc^{k}$}
\STATE{Set $\delta^{abs}_{\vp},\delta^{abs}_{\vc},\delta^{rel}_{\vp},\delta^{rel}_{\vc}\in\mathbb{R}^+$}
\FOR{$l\ge 0$}
\STATE{Define $f_{\vp}(\vp) = J(\vp,\vc^{(k,l)}) +\frac{\rho}{2} \|\D_{\vp} \vp - \z_{\vp}^k + \frac{\vl_{\vp}^k}{\rho}\|^2_2$}
\STATE{$\vp^{(k,l+1)} = \argmin_{\vp} f(\vp) + I_{\X^{SB}_{\vp}} (\vp) $} \label{subsubp0}
\STATE{Define $f_{\vc}(\vc) = J(\vp^{(k,l+1)},\vc)  +\frac{\rho}{2} \|\D_{\vc} \vc - \z_{\vc}^k + \frac{\vl_{\vc}^k}{\rho}\|^2_2$}
\STATE{$\vc^{(k,l+1)} = \argmin_{\vc} f(\vc) + I_{\X^{SB}_{\vc}} (\vc)$} \label{subsubc}
\STATE{$g_{\vp} = \|\vp^{(k,l+1)} - \vp^{(k,l)}\|_{\infty} - \delta^{rel}_{\vp}\|\vp^{(k,l+1)} -\vp^{w}\|_{\infty}$}
\STATE{$g_{\vc} = \|\vc^{(k,l+1)} - \vc^{(k,l)}\|_{\infty} - \delta^{rel}_{\vc}\|\vc^{(k,l+1)} -\vc^{w}\|_{\infty}$}
\IF{$g_{\vp} \le\delta^{abs}_{\vp}$ and $ g_{\vc}\le\delta^{abs}_{\vc}$}
\STATE{\textbf{return} $\x^{k+1} = \left[\vp^{(k,l+1)}; \vc^{(k,l+1)}\right]$}
\ENDIF

\ENDFOR
\end{algorithmic}
\end{algorithm}


\begin{algorithm}[!!htp]
\caption{Nonmonotone BB-based projected gradient method to solve the two sub-problems in Algorithm \ref{alg:step 1}}
\label{alg:sub}
\begin{algorithmic}[1]
\STATE{Denote $\vf$ as either $\vp$ or $\vc$}
\STATE{Initialize $\beta_{\vf}^0\in\mathbb{R}^+$, $\vf^{(k,l,0)} \gets \vf^{(k,l)}$}
\STATE{Set $\eta_{\vf}^{abs}, \eta_{\vf}^{rel}\in\mathbb{R}^+$,  $\gamma=10^{-4}$, $Q=10$, $s=0.5$}
\FOR{$j\ge 0$}
\STATE{$\vec{g}^j=\nabla_{\vf}f_{\vf}(\vf) \big|_{\vf=\vf^{(k,l,j)}}$}
\STATE{$\beta_{\vf}^j = \frac{\langle\vec{s}^{j-1},\vec{s}^{j-1}\rangle}{\langle\vec{s}^{j-1},\vec{y}^{j-1}\rangle}$ for $j\ge 1$, where $\vec{s}^{j-1}=\vf^{(k,l,j)}- \vf^{(k,l,j-1)}$ and $\vec{y}^{j-1}=\vec{g}^{j}- \vec{g}^{j-1}$}
\STATE{$\vec{h}^j=P_{I_{\X^{SB}_{\vf}}}\left(\vf^{(k,l,j)} -\beta_{\vf}^j\vec{g}^j\right) -\vf^{(k,l,j)} $}
\STATE{Set $\mu\leftarrow 1$}
\WHILE{True} 
    \STATE{Compute $\vf' = \vf^{(k,l,j)} + \mu \vec{h}^j$}
    \IF{$f_{\vf}(\vf') \le \max_{0\le q\le \min\{j,Q-1\}}f(\vf^{(k,l,j-q)}) +\gamma \mu\langle\vec{h}^j,\vec{g}^j\rangle$}
        \STATE{ $\vf^{(k,l,j+1)}\gets\vf'$}
        \STATE{\textbf{return } $\vf^{(k,l,j+1)}$}
    \ELSE
        \STATE{$\mu\leftarrow s\cdot\mu$}
    \ENDIF
\ENDWHILE
\IF{$\|\vf^{(k,l,j+1)} - \vf^{(k,l,j)}\|_{\infty}\le \eta^{rel}_{\vf} \|\vf^{(k,l,j+1)}-\vf^{w}\|_{\infty} +\eta^{abs}_{\vf}$}
    \STATE{\textbf{return } $\vf^{(k,l,j+1)}$}
\ENDIF
\ENDFOR
\end{algorithmic}
\end{algorithm}

\subsubsection{$\z$-minimization step.}
In  line \ref{subz} in Algorithm \ref{alg:admm}, $\z_{\vp}$ and $\z_{\vc}$ can be separately optimized as they are independent. 
For $\vf$, which denotes either $\vp$ or $\vc$, the projected least square solution $\z_{\vf}^{k+1}$ is obtained as 
\begin{equation}\label{eq:z1}
    \z_{\vf}^{k+1} = \mathrm{Proj}_{I_{\Z_{\vf}}} \left(\D_{\vf}\vf^{k+1} + \frac{\vl^{k}_{\vf}}{\rho}\right),
\end{equation} 
where $\mathrm{Proj}_{I_{\Z_{\vf}}}(\z_{\vf})$ is evaluated by solving the constrained convex problem \cite{liu2012multi}:
\begin{equation}
    \mathrm{Proj}_{I_{\Z_{\vf}}}(\z_{\vf}) = \argmin_{\z'_{\vf}\in \Z_{\vf}} \frac{1}{2}\| \z'_{\vf} - \z_{\vf}\|^2_2.
\end{equation}
Here, $\Z_{\vf} = \{\z= \left[\z_{\vf}^x; \z_{\vf}^y\right] \mid \|\z_{\vf}\|_{2,1} \le \tau_{\vf}, \z_i \in \mathbb{R}^N \}$ defines the $\ell_{2,1}$ ball with the radius of $\tau_{\vf}$.

\subsubsection{Optimization parameters settings.} 
\label{App:opt}
\begin{figure}[tbhp]
  \centering
     \subfloat[]{\label{fig:ac}\includegraphics[width=0.48\textwidth]{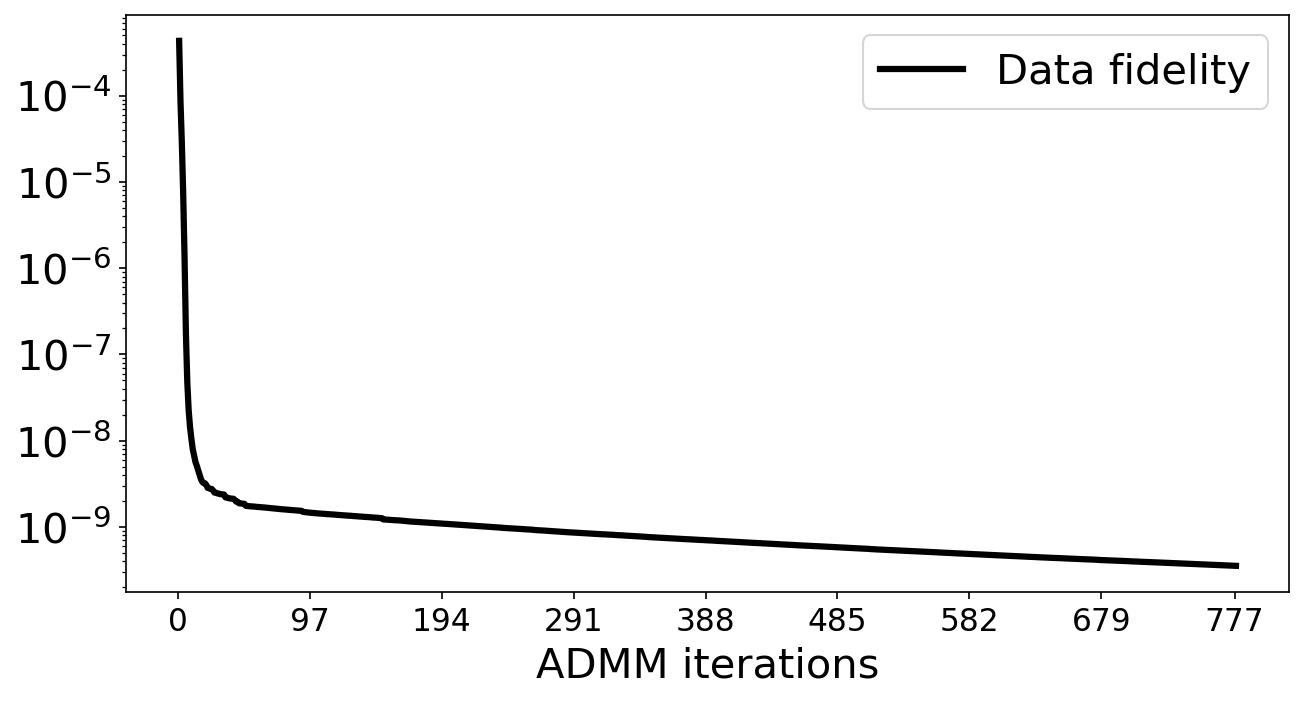}}
     \subfloat[]{\label{fig:bc}\includegraphics[width=0.48\textwidth]{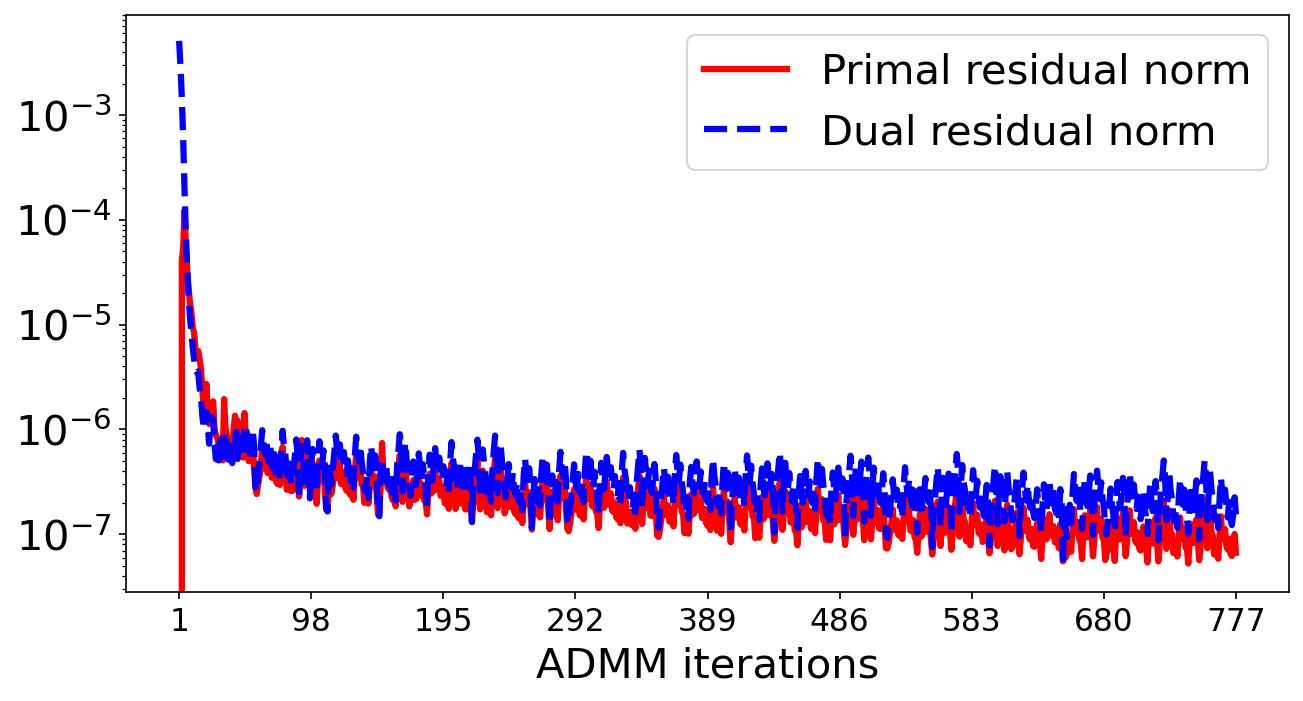}}
\caption{Convergence of (a) the data fidelity term $J(\x)$ as a function of ADMM iterations and (b) the primal and dual residual norms across ADMM iterations. The case corrsponds to NBP 1 (D-type) in \textbf{Study I}.}
  \label{fig:conv}
\end{figure}

The optimization parameters used across all numerical studies are summarized in Table \ref{tab:optimization}. 
A convergence analysis was performed using this parameter configuration, with the results shown in Fig.\ \ref{fig:conv} for the NBP 1 (D-type) example from \textbf{Study I}, shown in Fig.\ \ref{fig:study1}. 
Since this case is a strict inverse crime, the data fidelity term (Fig. \ref{fig:ac}) exhibited steady and smooth convergence, gradually approaching a minimum over the iterations. 
Additionally, the primal and dual residual norms (Fig. \ref{fig:bc}) also decreased steadily, though not strictly monotonically, with convergence rates balanced through adaptive tuning of the parameter $\rho$.

\begin{table}[!htp]
\footnotesize
\caption{Optimization parameters for the cases where TV constraints are applied}\label{tab:optimization}
\begin{center}
  \begin{tabular}{|c|c|c|} \hline 
    & \textbf{Study I} & \textbf{Studies II, III $\&$ IV} \\ \hline  
    \multirow{2}{*}{Algorithm \ref{alg:admm}}
    & Initial $\rho$=1, $\z^0$=$\vz$, & Initial $\rho$=1, $\z^0$=$\vz$ \\
    & $\epsilon^{abs}$=$2\cdot10^{-11}$, $\epsilon^{rel}$=$2\cdot10^{-12}$ & $\epsilon^{abs}$=$10^{-10}$, $\epsilon^{rel}$=$10^{-11}$ \\\hline
    \multirow{2}{*}{Algorithm \ref{alg:step 1}}&
       $\delta^{abs}_{\vp}$=$5\cdot10^{-7}$, $\delta^{rel}_{\vp}$=$10^{-7}$ & $\delta^{abs}_{\vp}$=$10^{-5}$, $\delta^{rel}_{\vp}$=$10^{-5}$ \\ &
    $\delta^{abs}_{\vc}$=$5\cdot10^{-5}$, $\delta^{rel}_{\vc}$=$10^{-7}$ & $\delta^{abs}_{\vc}$=$10^{-3}$, $\delta^{rel}_{\vc}$=$10^{-5}$ \\\hline
    \multirow{3}{*}{Algorithm \ref{alg:sub}}&$\eta^{abs}_{\vp}$=$5\cdot10^{-7}$, $\eta^{rel}_{\vp}$=$10^{-7}$ & $\eta^{abs}_{\vp}$=$10^{-6}$, $\eta^{rel}_{\vp}$=$10^{-6}$ \\
    &$\eta^{abs}_{\vc}$=$5\cdot 10^{-5}$, $\eta^{rel}_{\vc}$=$10^{-7}$ &$\delta^{abs}_{\vc}$=$10^{-3}$, $\delta^{rel}_{\vc}$=$10^{-5}$ \\
    &$\beta^0_{\vp}=2^2$, $\beta^0_{\vc}=2^{18}$&$\beta^0_{\vp}=2^2$, $\beta^0_{\vc}=2^{18}$ \\\hline
  \end{tabular}
\end{center}
\end{table}

\section*{References}
\bibliographystyle{plain}
\bibliography{references} 

\begin{thebibliography}{10}

\bibitem{attouch2010proximal}
H{\'e}dy Attouch, J{\'e}r{\^o}me Bolte, Patrick Redont, and Antoine Soubeyran.
\newblock Proximal alternating minimization and projection methods for nonconvex problems: An approach based on the kurdyka-{\l}ojasiewicz inequality.
\newblock {\em Mathematics of operations research}, 35(2):438--457, 2010.

\bibitem{barzilai1988two}
Jonathan Barzilai and Jonathan~M Borwein.
\newblock Two-point step size gradient methods.
\newblock {\em IMA journal of numerical analysis}, 8(1):141--148, 1988.

\bibitem{berenger1994perfectly}
Jean-Pierre Berenger.
\newblock A perfectly matched layer for the absorption of electromagnetic waves.
\newblock {\em Journal of computational physics}, 114(2):185--200, 1994.

\bibitem{bertsekas2014constrained}
Dimitri~P Bertsekas.
\newblock {\em Constrained optimization and Lagrange multiplier methods}.
\newblock Academic press, 2014.

\bibitem{birgin2000nonmonotone}
Ernesto~G Birgin, Jos{\'e}~Mario Mart{\'\i}nez, and Marcos Raydan.
\newblock Nonmonotone spectral projected gradient methods on convex sets.
\newblock {\em SIAM Journal on Optimization}, 10(4):1196--1211, 2000.

\bibitem{boyd2011distributed}
Stephen Boyd, Neal Parikh, Eric Chu, Borja Peleato, Jonathan Eckstein, et~al.
\newblock Distributed optimization and statistical learning via the alternating direction method of multipliers.
\newblock {\em Foundations and Trends{\textregistered} in Machine learning}, 3(1):1--122, 2011.

\bibitem{jax2018github}
James Bradbury, Roy Frostig, Peter Hawkins, Matthew~James Johnson, Chris Leary, Dougal Maclaurin, George Necula, Adam Paszke, Jake Vander{P}las, Skye Wanderman-{M}ilne, and Qiao Zhang.
\newblock {JAX}: composable transformations of {P}ython+{N}um{P}y programs, 2018.

\bibitem{cai2019feature}
Chuangjian Cai, Xuanhao Wang, Ke~Si, Jun Qian, Jianwen Luo, and Cheng Ma.
\newblock Feature coupling photoacoustic computed tomography for joint reconstruction of initial pressure and sound speed in vivo.
\newblock {\em Biomedical optics express}, 10(7):3447--3462, 2019.

\bibitem{chen2012non}
Xiaojun Chen, Michael~K Ng, and Chao Zhang.
\newblock Non-lipschitz $\ell_p$-regularization and box constrained model for image restoration.
\newblock {\em IEEE Transactions on Image Processing}, 21(12):4709--4721, 2012.

\bibitem{cong2015photoacoustic}
Bing Cong, Kengo Kondo, Takeshi Namita, Makoto Yamakawa, and Tsuyoshi Shiina.
\newblock Photoacoustic image quality enhancement by estimating mean sound speed based on optimum focusing.
\newblock {\em Japanese Journal of Applied Physics}, 54(7S1):07HC13, 2015.

\bibitem{dogan2019optoacoustic}
Basak~E Dogan, Gisela~LG Menezes, Reni~S Butler, Erin~I Neuschler, Roger Aitchison, Philip~T Lavin, F~Lee Tucker, Stephen~R Grobmyer, Pamela~M Otto, and A~Thomas Stavros.
\newblock Optoacoustic imaging and gray-scale us features of breast cancers: correlation with molecular subtypes.
\newblock {\em Radiology}, 292(3):564--572, 2019.

\bibitem{ermilov2009development}
Sergey~A Ermilov, Matthew~P Fronheiser, Hans-Peter Brecht, Richard Su, Andr{\'e} Conjusteau, Ketan Mehta, Pamela Otto, and Alexander~A Oraevsky.
\newblock Development of laser optoacoustic and ultrasonic imaging system for breast cancer utilizing handheld array probes.
\newblock In {\em Photons Plus Ultrasound: Imaging and Sensing 2009}, volume 7177, page 717703. International Society for Optics and Photonics, 2009.

\bibitem{esser2018total}
Ernie Esser, Lluis Guasch, Tristan van Leeuwen, Aleksandr~Y Aravkin, and Felix~J Herrmann.
\newblock Total variation regularization strategies in full-waveform inversion.
\newblock {\em SIAM Journal on Imaging Sciences}, 11(1):376--406, 2018.

\bibitem{fang2009monte}
Qianqian Fang and David~A Boas.
\newblock {M}onte {C}arlo simulation of photon migration in 3{D} turbid media accelerated by graphics processing units.
\newblock {\em Optics express}, 17(22):20178--20190, 2009.

\bibitem{he2000alternating}
Bing-Sheng He, Hai Yang, and SL~Wang.
\newblock Alternating direction method with self-adaptive penalty parameters for monotone variational inequalities.
\newblock {\em Journal of Optimization Theory and applications}, 106:337--356, 2000.

\bibitem{herzfeld2013absorption}
Karl~F Herzfeld and Theodore~A Litovitz.
\newblock {\em Absorption and dispersion of ultrasonic waves}, volume~7.
\newblock Academic Press, 2013.

\bibitem{huang2010investigation}
Chao Huang, Alexander~A Oraevsky, and Mark~A Anastasio.
\newblock Investigation of limited-view image reconstruction in optoacoustic tomography employing a priori structural information.
\newblock In {\em Image Reconstruction from Incomplete Data VI}, volume 7800, pages 27--32. SPIE, 2010.

\bibitem{huang2013full}
Chao Huang, Kun Wang, Liming Nie, Lihong~V Wang, and Mark~A Anastasio.
\newblock Full-wave iterative image reconstruction in photoacoustic tomography with acoustically inhomogeneous media.
\newblock {\em IEEE transactions on medical imaging}, 32(6):1097--1110, 2013.

\bibitem{huang2016joint}
Chao Huang, Kun Wang, Robert~W Schoonover, Lihong~V Wang, and Mark~A Anastasio.
\newblock Joint reconstruction of absorbed optical energy density and sound speed distributions in photoacoustic computed tomography: a numerical investigation.
\newblock {\em IEEE transactions on computational imaging}, 2(2):136--149, 2016.

\bibitem{jiang2006spatially}
Huabei Jiang, Zhen Yuan, and Xuejun Gu.
\newblock Spatially varying optical and acoustic property reconstruction using finite-element-based photoacoustic tomography.
\newblock {\em JOSA A}, 23(4):878--888, 2006.

\bibitem{jin2022signal}
Gaofei Jin, Hui Zhu, Daohuai Jiang, Jinwei Li, Lili Su, Jianfeng Li, Fei Gao, and Xiran Cai.
\newblock A signal-domain object segmentation method for ultrasound and photoacoustic computed tomography.
\newblock {\em IEEE Transactions on Ultrasonics, Ferroelectrics, and Frequency Control}, 70(3):253--265, 2022.

\bibitem{jin2006thermoacoustic}
Xing Jin and Lihong~V Wang.
\newblock Thermoacoustic tomography with correction for acoustic speed variations.
\newblock {\em Physics in Medicine \& Biology}, 51(24):6437, 2006.

\bibitem{jose2012speed}
Jithin Jose, Rene~GH Willemink, Wiendelt Steenbergen, Cornelis~H Slump, Ton~G van Leeuwen, and Srirang Manohar.
\newblock Speed-of-sound compensated photoacoustic tomography for accurate imaging.
\newblock {\em Medical physics}, 39(12):7262--7271, 2012.

\bibitem{kirsch2012simultaneous}
Andreas Kirsch and Otmar Scherzer.
\newblock Simultaneous reconstructions of absorption density and wave speed with photoacoustic measurements.
\newblock {\em SIAM Journal on Applied Mathematics}, 72(5):1508--1523, 2012.

\bibitem{kruger2013dedicated}
Robert~A Kruger, Cherie~M Kuzmiak, Richard~B Lam, Daniel~R Reinecke, Stephen~P Del~Rio, and Doreen Steed.
\newblock Dedicated 3{D} photoacoustic breast imaging.
\newblock {\em Medical physics}, 40(11):113301, 2013.

\bibitem{kruger2010photoacoustic}
Robert~A Kruger, Richard~B Lam, Daniel~R Reinecke, Stephen~P Del~Rio, and Ryan~P Doyle.
\newblock Photoacoustic angiography of the breast.
\newblock {\em Medical physics}, 37(11):6096--6100, 2010.

\bibitem{li20213}
Fu~Li, Umberto Villa, Seonyeong Park, and Mark~A Anastasio.
\newblock 3-{D} stochastic numerical breast phantoms for enabling virtual imaging trials of ultrasound computed tomography.
\newblock {\em IEEE transactions on ultrasonics, ferroelectrics, and frequency control}, 69(1):135--146, 2021.

\bibitem{li2017single}
Lei Li, Liren Zhu, Cheng Ma, Li~Lin, Junjie Yao, Lidai Wang, Konstantin Maslov, Ruiying Zhang, Wanyi Chen, Junhui Shi, et~al.
\newblock Single-impulse panoramic photoacoustic computed tomography of small-animal whole-body dynamics at high spatiotemporal resolution.
\newblock {\em Nature biomedical engineering}, 1(5):0071, 2017.

\bibitem{lin2018single}
Li~Lin, Peng Hu, Junhui Shi, Catherine~M Appleton, Konstantin Maslov, Lei Li, Ruiying Zhang, and Lihong~V Wang.
\newblock Single-breath-hold photoacoustic computed tomography of the breast.
\newblock {\em Nature communications}, 9(1):2352, 2018.

\bibitem{liu2015determining}
Hongyu Liu and Gunther Uhlmann.
\newblock Determining both sound speed and internal source in thermo-and photo-acoustic tomography.
\newblock {\em Inverse Problems}, 31(10):105005, 2015.

\bibitem{liu2012multi}
Jun Liu, Shuiwang Ji, and Jieping Ye.
\newblock Multi-task feature learning via efficient l2, 1-norm minimization.
\newblock {\em arXiv preprint arXiv:1205.2631}, 2012.

\bibitem{mandal2016visual}
Subhamoy Mandal, Xos{\'e}~Lu{\'\i}s De{\'a}n-Ben, and Daniel Razansky.
\newblock Visual quality enhancement in optoacoustic tomography using active contour segmentation priors.
\newblock {\em IEEE transactions on medical imaging}, 35(10):2209--2217, 2016.

\bibitem{manohar2007concomitant}
Srirang Manohar, Ren{\'e}~GH Willemink, Ferdi van~der Heijden, Cornelis~H Slump, and Ton~G van Leeuwen.
\newblock Concomitant speed-of-sound tomography in photoacoustic imaging.
\newblock {\em Applied physics letters}, 91(13), 2007.

\bibitem{matthews2017joint}
Thomas~P Matthews and Mark~A Anastasio.
\newblock Joint reconstruction of the initial pressure and speed of sound distributions from combined photoacoustic and ultrasound tomography measurements.
\newblock {\em Inverse problems}, 33(12):124002, 2017.

\bibitem{matthews2018parameterized}
Thomas~P Matthews, Joemini Poudel, Lei Li, Lihong~V Wang, and Mark~A Anastasio.
\newblock Parameterized joint reconstruction of the initial pressure and sound speed distributions for photoacoustic computed tomography.
\newblock {\em SIAM journal on imaging sciences}, 11(2):1560--1588, 2018.

\bibitem{modgil2010image}
Dimple Modgil, Mark~A Anastasio, and Patrick~J La~Rivi{\`e}re.
\newblock Image reconstruction in photoacoustic tomography with variable speed of sound using a higher-order geometrical acoustics approximation.
\newblock {\em Journal of biomedical optics}, 15(2):021308--021308, 2010.

\bibitem{nocedal1999numerical}
Jorge Nocedal and Stephen~J Wright.
\newblock {\em Numerical optimization}.
\newblock Springer, 1999.

\bibitem{oksanen2013photoacoustic}
Lauri Oksanen and Gunther Uhlmann.
\newblock Photoacoustic and thermoacoustic tomography with an uncertain wave speed.
\newblock {\em arXiv preprint arXiv:1307.1618}, 2013.

\bibitem{parikh2014proximal}
Neal Parikh, Stephen Boyd, et~al.
\newblock Proximal algorithms.
\newblock {\em Foundations and trends{\textregistered} in Optimization}, 1(3):127--239, 2014.

\bibitem{park2023stochastic}
Seonyeong Park, Umberto Villa, Fu~Li, Refik~Mert Cam, Alexander~A Oraevsky, and Mark~A Anastasio.
\newblock Stochastic three-dimensional numerical phantoms to enable computational studies in quantitative optoacoustic computed tomography of breast cancer.
\newblock {\em Journal of Biomedical Optics}, 28(6):066002--066002, 2023.

\bibitem{parker2022power}
KJ~Parker.
\newblock Power laws prevail in ultrasound-tissue interactions.
\newblock {\em Physics in medicine and biology}, 67(9), 2022.

\bibitem{poudel2019survey}
Joemini Poudel, Yang Lou, and Mark~A Anastasio.
\newblock A survey of computational frameworks for solving the acoustic inverse problem in three-dimensional photoacoustic computed tomography.
\newblock {\em Physics in Medicine \& Biology}, 64(14):14TR01, 2019.

\bibitem{radiology2013acr}
ACo Radiology and CJ~D'Orsi.
\newblock {ACR BI-RADS} atlas: breast imaging reporting and data system; mammography, ultrasound, magnetic resonance imaging, follow-up and outcome monitoring, data dictionary.
\newblock {\em ACR, American College of Radiology}, 2013.

\bibitem{ranjbaran2023quantitative}
Seyed~Mohsen Ranjbaran, Hossein~S Aghamiry, Ali Gholami, St{\'e}phane Operto, and Kamran Avanaki.
\newblock Quantitative photoacoustic tomography using iteratively refined wavefield reconstruction inversion: a simulation study.
\newblock {\em IEEE Transactions on Medical Imaging}, 2023.

\bibitem{ren2024simultaneous}
Zhimei Ren, Emil~Y Sidky, Rina~Foygel Barber, Chien-Min Kao, and Xiaochuan Pan.
\newblock Simultaneous activity and attenuation estimation in {TOF-PET} with {TV}-constrained nonconvex optimization.
\newblock {\em IEEE Transactions on Medical Imaging}, 2024.

\bibitem{sidky2012convex}
Emil~Y Sidky, Jakob~H J{\o}rgensen, and Xiaochuan Pan.
\newblock Convex optimization problem prototyping for image reconstruction in computed tomography with the {C}hambolle--{P}ock algorithm.
\newblock {\em Physics in Medicine \& Biology}, 57(10):3065, 2012.

\bibitem{stanziola2023j}
Antonio Stanziola, Simon~R Arridge, Ben~T Cox, and Bradley~E Treeby.
\newblock j-{W}ave: An open-source differentiable wave simulator.
\newblock {\em SoftwareX}, 22:101338, 2023.

\bibitem{stefanov2012instability}
Plamen Stefanov and Gunther Uhlmann.
\newblock Instability of the linearized problem in multiwave tomography of recovery both the source and the speed.
\newblock {\em arXiv preprint arXiv:1211.6217}, 2012.

\bibitem{tabei2002k}
Makoto Tabei, T~Douglas Mast, and Robert~C Waag.
\newblock A k-space method for coupled first-order acoustic propagation equations.
\newblock {\em The Journal of the Acoustical Society of America}, 111(1):53--63, 2002.

\bibitem{treeby2010modeling}
Bradley~E Treeby and Ben~T Cox.
\newblock Modeling power law absorption and dispersion for acoustic propagation using the fractional laplacian.
\newblock {\em The Journal of the Acoustical Society of America}, 127(5):2741--2748, 2010.

\bibitem{treeby2010photoacoustic}
Bradley~E Treeby, Edward~Z Zhang, and Benjamin~T Cox.
\newblock Photoacoustic tomography in absorbing acoustic media using time reversal.
\newblock {\em Inverse Problems}, 26(11):115003, 2010.

\bibitem{vaswani2010modified}
Namrata Vaswani and Wei Lu.
\newblock Modified-{CS}: Modifying compressive sensing for problems with partially known support.
\newblock {\em IEEE Transactions on Signal Processing}, 58(9):4595--4607, 2010.

\bibitem{wang2012simple}
Kun Wang and Mark~A Anastasio.
\newblock A simple fourier transform-based reconstruction formula for photoacoustic computed tomography with a circular or spherical measurement geometry.
\newblock {\em Physics in Medicine \& Biology}, 57(23):N493, 2012.

\bibitem{wang2017photoacoustic}
Lihong~V Wang.
\newblock {\em Photoacoustic imaging and spectroscopy}.
\newblock CRC press, 2017.

\bibitem{wang2012biomedical}
Lihong~V Wang and Hsin-i Wu.
\newblock {\em Biomedical optics: principles and imaging}.
\newblock John Wiley \& Sons, 2012.

\bibitem{wang2001decomposition}
SL~Wang and LZ~Liao.
\newblock Decomposition method with a variable parameter for a class of monotone variational inequality problems.
\newblock {\em Journal of optimization theory and applications}, 109:415--429, 2001.

\bibitem{wang2010sparse}
Yilun Wang and Wotao Yin.
\newblock Sparse signal reconstruction via iterative support detection.
\newblock {\em SIAM Journal on Imaging Sciences}, 3(3):462--491, 2010.

\bibitem{xia2012whole}
Jun Xia, Muhammad~R Chatni, Konstantin Maslov, Zijian Guo, Kun Wang, Mark Anastasio, and Lihong~V Wang.
\newblock Whole-body ring-shaped confocal photoacoustic computed tomography of small animals in vivo.
\newblock {\em Journal of biomedical optics}, 17(5):050506--050506, 2012.

\bibitem{xia2013enhancement}
Jun Xia, Chao Huang, Konstantin Maslov, Mark~A Anastasio, and Lihong~V Wang.
\newblock Enhancement of photoacoustic tomography by ultrasonic computed tomography based on optical excitation of elements of a full-ring transducer array.
\newblock {\em Optics letters}, 38(16):3140--3143, 2013.

\bibitem{xu2005universal}
Minghua Xu and Lihong~V Wang.
\newblock Universal back-projection algorithm for photoacoustic computed tomography.
\newblock {\em Physical Review E}, 71(1):016706, 2005.

\bibitem{xu2003time}
Minghua Xu, Yuan Xu, and Lihong~V Wang.
\newblock Time-domain reconstruction algorithms and numerical simulations for thermoacoustic tomography in various geometries.
\newblock {\em IEEE Transactions on biomedical engineering}, 50(9):1086--1099, 2003.

\bibitem{xu2003effects}
Yuan Xu and Lihong~V Wang.
\newblock Effects of acoustic heterogeneity in breast thermoacoustic tomography.
\newblock {\em IEEE transactions on ultrasonics, ferroelectrics, and frequency control}, 50(9):1134--1146, 2003.

\bibitem{yoon2012enhancement}
Changhan Yoon, Jeeun Kang, Seunghee Han, Yangmo Yoo, Tai-Kyong Song, and Jin~Ho Chang.
\newblock Enhancement of photoacoustic image quality by sound speed correction: ex vivo evaluation.
\newblock {\em Optics express}, 20(3):3082--3090, 2012.

\bibitem{yu2018scalable}
Leiming Yu, Fanny Nina-Paravecino, David Kaeli, and Qianqian Fang.
\newblock Scalable and massively parallel {M}onte {C}arlo photon transport simulations for heterogeneous computing platforms.
\newblock {\em Journal of biomedical optics}, 23(1):010504--010504, 2018.

\bibitem{yuan2006simultaneous}
Zhen Yuan, Qizhi Zhang, and Huabei Jiang.
\newblock Simultaneous reconstruction of acoustic and optical properties of heterogeneous media by quantitative photoacoustic tomography.
\newblock {\em Optics express}, 14(15):6749--6754, 2006.

\bibitem{zhang2006reconstruction}
Jin Zhang and Mark~A Anastasio.
\newblock Reconstruction of speed-of-sound and electromagnetic absorption distributions in photoacoustic tomography.
\newblock In {\em Photons Plus Ultrasound: Imaging and Sensing 2006: The Seventh Conference on Biomedical Thermoacoustics, Optoacoustics, and Acousto-optics}, volume 6086, pages 339--345. SPIE, 2006.

\bibitem{zhang2008simultaneous}
Jin Zhang, Kun Wang, Yongyi Yang, and Mark~A Anastasio.
\newblock Simultaneous reconstruction of speed-of-sound and optical absorption properties in photoacoustic tomography via a time-domain iterative algorithm.
\newblock In {\em Photons Plus Ultrasound: Imaging and Sensing 2008: The Ninth Conference on Biomedical Thermoacoustics, Optoacoustics, and Acousto-optics}, volume 6856, page 68561F. International Society for Optics and Photonics, 2008.

\bibitem{zhang2016investigation}
Zheng Zhang, Jinghan Ye, Buxin Chen, Amy~E Perkins, Sean Rose, Emil~Y Sidky, Chien-Min Kao, Dan Xia, Chi-Hua Tung, and Xiaochuan Pan.
\newblock Investigation of optimization-based reconstruction with an image-total-variation constraint in {PET}.
\newblock {\em Physics in Medicine \& Biology}, 61(16):6055, 2016.

\end{thebibliography}
\end{document}